\documentclass{aa}

\usepackage{graphicx}
\usepackage{natbib}
\usepackage{lscape}
\usepackage{longtable}
\usepackage{color}
\usepackage{txfonts}
\begin{document}

   \title{The HADES RV Programme with HARPS-N@TNG
   \thanks{
     Based on observations made with the Italian Telescopio Nazionale Galileo (TNG), operated
     on the island of La Palma by the Fundaci\'on Galileo Galilei of the INAF
     (Istituto Nazionale di Astrofisica) at the Spanish Observatorio del Roque de
     los Muchachos of the Instituto de Astrof\'isica de Canarias. 
     }\fnmsep
   \thanks{
     Tables~\ref{parameters_table_full}, ~\ref{kinematic_catalogue},
     and ~\ref{parameters_emission_excesses} are only available 
     in the electronic version of the paper or at the CDS via anonymous
     ftp to cdsarc.u-strasbg.fr (130.79.128.5)
     or via http://cdsweb.u-strasbg.fr/cgi-bin/qcat?J/A+A/
     }
     }

   \subtitle{III. Flux-flux and activity-rotation relationships of early-M dwarfs
         }

   \author{J. Maldonado 
          \inst{1}
          \and G. Scandariato 
          \inst{2} 
          \and  B. Stelzer
          \inst{1}
          \and  K. Biazzo
          \inst{2}
          \and  A.F. Lanza
          \inst{2}
          \and A. Maggio
          \inst{1}   
          \and G. Micela
          \inst{1} 
          \and E. Gonz\'alez -\'Alvarez
          \inst{3,1}
          \and L. Affer
          \inst{1} 
          \and R. U. Claudi
          \inst{4}
          \and R. Cosentino
          \inst{2,5}   
          \and M. Damasso
          \inst{6} 
          \and S. Desidera
          \inst{4}  
          \and J. I. Gonz\'alez Hern\'andez
          \inst{7,8}
          \and R. Gratton
          \inst{4}
          \and G. Leto
          \inst{2}
          \and S. Messina
          \inst{2}
          \and E. Molinari
          \inst{5,9}
          \and I. Pagano  
          \inst{2}
          \and M. Perger
          \inst{10}
          \and G. Piotto
          \inst{11,4}
          \and R. Rebolo
          \inst{7,8} 
          \and I. Ribas
          \inst{10}
          \and A. Sozzetti
          \inst{6}
          \and A. Su\'arez Mascare\~no
          \inst{7,8}
          \and R. Zanmar Sanchez
          \inst{2}
          } 

             \institute{INAF - Osservatorio Astronomico di Palermo,
              Piazza del Parlamento 1, 90134 Palermo, Italy
             \and 
              INAF - Osservatorio Astrofisico di Catania, Via S. Sofia 78, 95123, Catania, Italy
             \and 
              Dipartimento di Fisica \& Chimica, Universit\`a di Palermo, Piazza del Parlamento 1, 90134 Palermo, Italy  
             \and 
              INAF - Osservatorio Astronomico di Padova, Vicolo Osservatorio 5, 35122 Padova, Italy
             \and 
              Fundaci\'on Galileo Galilei - INAF, Rambla Jos\'e Ana Fernandez P\'erez 7, 38712 Brea\~na Baja, TF, Spain
             \and 
              INAF - Osservatorio Astrofisico di Torino, Via Osservatorio 20, 10025, Pino Torinese, Italy
             \and  
              Instituto de Astrof\'isica de Canarias, 38205 La Laguna, Tenerife, Spain
             \and  
              Universidad de La Laguna, Dpto. Astrof\'isica, 38206 La Laguna, Tenerife, Spain
             \and 
              INAF - IASF Milano, via Bassini 15, 20133 Milano, Italy 
             \and  
              Instittut de Ci\`encies de l'Espai (IEEC-CSIC), Campus UAB, C/ Can Magrans, s/n, 08193 Bellaterra, Spain
             \and 
             Dip. di Fisica e Astronomia Galileo Galilei – Universit\`a di Padova,
             Vicolo dell'Osservatorio 2, 35122, Padova, Italy
             }

   \offprints{J. Maldonado \\ \email{jmaldonado@astropa.inaf.it}}
   \date{Received ...; Accepted ....}

 
  \abstract
   {
    Understanding stellar activity in M dwarfs is crucial for
    the physics of stellar atmospheres as well as for
    ongoing radial velocity exoplanet programmes.
    Despite the increasing interest in M dwarfs,  
    our knowledge of the chromospheres of these stars is far from being complete.  
    }
   {We aim to test whether the relations between activity, rotation, and stellar parameters and flux-flux relationships
    previously investigated for main-sequence FGK stars and for pre-main sequence M stars
    also hold for early-M dwarfs on the main-sequence.
    Although several attempts have been made so far, here we analyse a large
    sample of relatively low-activity stars. 
   }
   {
    We analyse in an homogeneous and coherent way a well defined sample of
    71 late-K/early-M dwarfs that are
    currently being observed in the framework of the HArps-n red Dwarf Exoplanet Survey (HADES).    
    Rotational velocities are derived using the cross-correlation technique while
    emission flux excesses in the Ca~{\sc ii} H \& K and Balmer lines from H$\alpha$ up to H$\epsilon$ are obtained by using
    the spectral subtraction technique. 
    The relationships between the emission excesses and the stellar parameters
    (projected rotational velocity, effective temperature, kinematics, and age) are studied. 
    Relations between pairs of fluxes of different chromospheric lines (flux-flux 
    relationships) are also studied and compared with the
    literature results for other samples of stars.
    }  
   {We find that the strength of the chromospheric emission in the Ca~{\sc ii} H \& K and Balmer lines 
    is roughly constant for stars in the M0-M3 spectral range. Although our sample is
    likely to be biased towards inactive stars, our data suggest that a moderate
    but significant correlation between activity and rotation  might be present as well as a hint of 
     kinematically selected young
    stars showing higher levels of emission in the Calcium and most of the Balmer lines. 
    We find our sample of M dwarfs to be complementary in terms of chromospheric and X-ray fluxes
    with those of the literature, extending the analysis of the flux-flux relationships to the very low flux
    domain. 
   }
   {Our results agree with previous works suggesting that the activity-rotation-age
   relationship known to hold for solar-type stars also applies to early-M dwarfs.
   We also 
   confirm previous findings that the field stars which deviate from the bulk of the empirical
   flux-flux relationships show evidence of youth.}


  \maketitle

\section{Introduction}
\label{introduccion}

 The outer atmospheres of cool stars show diverse types of 
 non-radiative heating associated with magnetic fields,
 a phenomenon globally known as ``activity''. 
 Non-radiative heating produced by acoustic waves 
 is responsible for the basal emission observed in
 inactive stars. 
 It is well known that
 in solar-type stars with convective outer layers,
 chromospheric activity and rotation are linked by the stellar dynamo
 \citep[e.g.][]{1967ApJ...150..551K,1984ApJ...279..763N,2001MNRAS.326..877M},
 and both activity and rotation diminish during the main-sequence phase as stars
 evolve \citep[e.g.][]{1972ApJ...171..565S,1989ApJ...343L..65K,1991ApJ...375..722S,2007ApJ...669.1167B,2008ApJ...687.1264M}.
 This is due to the loss of angular momentum via 
 magnetic braking \citep{1967ApJ...148..217W,1993MNRAS.261..766J}. 

 Activity is usually 
 observed in the cores of the Ca~{\sc ii} H \& K lines and the Balmer lines. Other common
 optical activity indicators include lines such as the Na D$_{\rm 1}$, D$_{\rm 2}$ doublet, 
 the Mg~{\sc i} b triplet, or the  Ca~{\sc ii} infrared triplet. 
 By performing a simultaneous analysis of different optical
 chromospheric activity indicators, a detailed study of the
 chromospheric structure can be carried out 
 \citep[e.g.][]{2000A&AS..146..103M,2001A&A...379..976M,2013A&A...558A.141S}.
 The common approach is to study the relationship between pairs of 
 fluxes of different lines. 
 After subtracting the contribution of the basal atmosphere 
 from the observed emission, the relationship between excess fluxes in 
 two different lines may be fitted by a power-law function
 \citep[e.g.][]{2000ssma.book.....S}.
 Since the pioneering works of \cite{1987A&A...172..111S} and \cite{1989A&A...219..239R},
 the relationship among different
 chromospheric indicators have been largely studied
 \citep[e.g.][]{1990ApJS...72..191S,1990ApJS...74..891R,1993MNRAS.262....1T,
 1995A&AS..114..287M,1996ASPC..109..657M,1996A&A...312..221M, 
 2005ESASP.560..775L,2007A&A...466.1089B,2007A&A...469..309C,2010A&A...520A..79M,2011MNRAS.414.2629M,
 2012A&A...537A..94S,2013A&A...558A.141S}.

 Low-mass, M dwarf stars, constitute (by number) the largest component
 of the solar neighbourhood, $\sim$
 75\% of the stars within 10 pc being M dwarfs\footnote{http://www.recons.org/census.posted.htm}
 \citep{2006AJ....132.2360H}. 
 However, the outer atmosphere of these stars
 remains poorly understood as their intrinsic faintness at optical wavelengths
 makes it difficult to obtain high-resolution data.
 Despite these difficulties, some studies suggest that the connection
 between age, rotation, and activity may also hold in 
 early-M dwarfs 
 \citep[e.g.][]{1998A&A...331..581D,2003A&A...410..671M,2003ApJ...583..451M,2003A&A...397..147P,2004AJ....128..426W,
 2007AcA....57..149K,2010AJ....139..504B,2012AJ....143...93R,2013MNRAS.431.2063S,2015ApJ...812....3W}; 
 although deviations in the case of close M binaries
 have been reported \citep{2014A&A...570A..19M}.


 Some previous works suggest that some M dwarfs may departure
 from the general flux-flux relationships in some spectral lines.
 \cite{1986A&A...154..185O,1987A&A...177..143S,1989A&A...219..239R}
 found deviation in soft X-rays and chromospheric and transition-region emission
 lines in M dwarfs with emission lines. In a later work,
 \cite{2005ESASP.560..775L} identified some deviating M dwarfs as possible 
 flare stars.
 More recently, \cite{2011MNRAS.414.2629M} performed a detailed analysis
 of the flux-flux relationships including a large sample of M stars. 
 In some correlations the authors identified two branches, an ``inactive'' one
 composed of field stars with spectral types from F to M, and a second one
 populated by a subsample of M field dwarfs. 
 They show that the deviating stars have saturated X-ray and H$\alpha$ emission,
 concluding that about 75\% of them have ages compatible with the Pleiades or
 younger.
 \cite{2013A&A...558A.141S} analysed a large set of emission lines for 
 a sample of 24 pre-main sequence M stars noting that all of them
 followed the ``active''-- branch defined by \cite{2011MNRAS.414.2629M}. 


 M dwarfs are nowadays becoming the main targets to search for 
 rocky, low-mass planets with the potential capability of hosting life
 \citep[e.g.][]{2013ApJ...767...95D,2013EPJWC..4703006S}.
 Understanding the chromospheres of M dwarfs is crucial for this purpose.
 Stellar activity, including stellar spots, as well as oscillations
 and granulation are challenging the detection of low-mass planets
 via radial velocity and transit surveys 
 \citep[e.g.][]{2012Natur.491..207D,2014prpl.conf..715F,2016A&A...586A.131H}.
 Further, the high-levels of activity (strong flares  cf. \citet{1997A&A...327.1114L,2005ApJ...621..398O}
 - and high UV emission in quiescence)
 of M dwarfs may constitute a potential hazard for habitability
 \citep{2013ApJ...763..149F}.

 In this paper we present a study of the activity-rotation-stellar parameters
 and flux-flux relationships for a large sample of early-M dwarfs
 that are currently being monitored in radial velocity surveys.
 In this paper we focus on average trends while the short-term chromospheric
 variability of the sample is studied
 in a companion paper \citet[][submitted]{Gaetano} 
 This paper is organised as follows. 
 Sect.~\ref{spectroscopic_data}
 describes the stellar sample and the spectroscopic
 data. The technique developed for determining rotational velocities
 is described in Sect.~\ref{rotational_velocities}.
 The analysis of the different activity indicators
 (Ca~{\sc ii} H \& K, and Balmer lines)
 is discussed
 in Sect.~\ref{spectral_subtraction}.
 Additional data (kinematics, X-ray emission) are presented
 and discussed in Sect.~\ref{spectral_others}.
 Results are given  and discussed in Sect.~\ref{results}. 
 Our conclusions follow in Sect.~\ref{summary}.

\section{Stellar sample}
\label{spectroscopic_data}

 Our stellar sample is composed of 78 late-K/early-M dwarfs
 monitored within the 
 HArps-n red Dwarf Exoplanet Survey (HADES),
 \citet[][]{Affer}; \citet[][submitted]{Perger},
 a collaborative effort between the 
 Global Architecture of Planetary Systems project \citep[GAPS;][]{2013A&A...554A..28C}
 \footnote{http://www.oact.inaf.it/exoit/EXO-IT/Projects/Entries/2011/12/\\27\_GAPS.html},
 the Institut de Ci\`encies de l’Espai (ICE/CSIC), and the Instituto de Astrof\'isica
 de Canarias (IAC).
 71 stars have been observed so far
 covering a range in effective temperature from 3400 to 3900 K, and 
 have spectral types between K7.5 and M3V.
 They were selected from the Palomar-Michigan State
 University (PMSU) catalogue \citep{1995AJ....110.1838R}, \cite{2011AJ....142..138L},
 and are targets observed within the 
 APACHE transit survey \citep{2013EPJWC..4703006S} with a
 visible magnitude lower than 12 and with an expected high number of {\sc Gaia}
 mission scans. 
 Analogous to other samples selected for Doppler searches, our sample is likely to be biased
 including mostly stars with low rotation rate and activity level. 

 All the observed stars show emission in
 the cores of the Ca~{\sc ii} H \& K lines.
 It is common in the literature to classify M dwarfs as
 active or inactive according to whether the core of the H${\alpha}$
 line shows or not emission \citep[see e.g.][and references therein]{2012AJ....143...93R}.
 Only three of our targets match this criterion. 
 We note that this criterion has some caveats as other diagnostics, 
 e.g., the Ca~{\sc ii} H \& K lines or X-ray emission,
 have been shown to be more sensitive for tracing low activity levels than H${\alpha}$
 \citep{2008ApJ...677..593W,2013MNRAS.431.2063S}.

 High-resolution \'echelle spectra of the stars were obtained at
 La Palma observatory (Canary Islands, Spain) during several observing
 runs between September 2012 and February 2016 
 using the HARPS-N \citep{2012SPIE.8446E..1VC} instrument 
 at the Telescopio Nazionale Galileo (TNG). 
  HARPS-N spectra cover the wavelength
 range 383-693 nm with a resolving power of $R$ $\sim$ 115000.
 All spectra were automatically reduced using
 the Data Reduction Software  
 \citep[DRS V3.7,][]{2007A&A...468.1115L}.

 Roughly 65\% of the stars have more
 than 15 observations, 
 the median number of observations per star being 27.
 For stars with more than one observation, spectra were combined
 into one single spectrum following the procedure described in
 \citet[][submitted]{Gaetano}. In the following we shall refer (unless otherwise noticed)
 to the combined spectra. 
 Basic stellar parameters (effective temperature, spectral type,
 surface gravity, iron abundance, mass, radius, and luminosity) were computed using a methodology
 based on ratios of spectral features\footnote{http://www.astropa.inaf.it/\textasciitilde{}jmaldonado/}
 \citep{2015A&A...577A.132M}. 
 Our estimates agree reasonable well with some previous detailed analysis of stellar parameters
 \citep[e.g. GJ 15A,][]{2014ApJ...794...51H}.
 Our sample is presented in Table~\ref{parameters_table_full},
 available online. 

\onllongtab{
\begin{longtable}{lcccccccc}
\caption{
{
 Basic stellar parameters of the stellar sample.
 }
  }\label{parameters_table_full}\\
\hline
\hline
 Star   & T$_{\rm eff}$ &  Sp-Type   & [Fe/H]   &  M$_{\star}$     & R$_{\star}$  & $\log g$   &  $\log (L_{\star}/L_{\odot})$ & $v\sin i$ \\
        &   (K)         &            & (dex)    &  (M$_{\odot}$)   & (R$_{\odot}$)& (cgs)    &                            & (km s$^{\rm -1}$) \\ 
 (1)    &   (2)         &  (3)       & (4)      &   (5)            & (6)             &  (7)           & (8)           &  (9) \\            
\hline
\endfirsthead
\caption{Continued.} \\
\hline
 Star   & T$_{\rm eff}$ &  Sp-Type   & [Fe/H]   &  M$_{\star}$     & R$_{\star}$  & $\log g$   &  $\log (L_{\star}/L_{\odot})$ &  $v\sin i$   \\
        &   (K)         &            & (dex)    &  (M$_{\odot}$)   & (R$_{\odot}$)& (cgs)      &                   &  (km s$^{\rm -1}$) \\
 (1)    &   (2)         &  (3)       & (4)      &   (5)            & (6)             &  (7)           & (8)    &  (9) \\
\hline
\endhead
\hline
\endfoot
\hline
\endlastfoot
GJ 2	&	3713	$\pm$	68	&	M1	&	-0.14	$\pm$	0.09	&	0.51	$\pm$	0.05	&	0.49	$\pm$	0.05	&	4.76	$\pm$	0.04	&	-1.380	$\pm$	0.087	&		0.98	$\pm$	0.54	\\
GJ 3014	&	3695	$\pm$	69	&	M1.5	&	-0.19	$\pm$	0.09	&	0.48	$\pm$	0.05	&	0.47	$\pm$	0.05	&	4.79	$\pm$	0.04	&	-1.433	$\pm$	0.090	&	$<$	1.08			\\
GJ 16	&	3673	$\pm$	68	&	M1.5	&	-0.16	$\pm$	0.09	&	0.48	$\pm$	0.05	&	0.47	$\pm$	0.05	&	4.78	$\pm$	0.04	&	-1.441	$\pm$	0.090	&	$<$	1.02			\\
GJ 15A	&	3607	$\pm$	68	&	M1	&	-0.34	$\pm$	0.09	&	0.38	$\pm$	0.05	&	0.38	$\pm$	0.05	&	4.87	$\pm$	0.04	&	-1.655	$\pm$	0.112	&		1.09	$\pm$	0.79	\\
GJ 21	&	3746	$\pm$	68	&	M1	&	-0.12	$\pm$	0.09	&	0.53	$\pm$	0.05	&	0.52	$\pm$	0.05	&	4.74	$\pm$	0.04	&	-1.328	$\pm$	0.086	&		1.46	$\pm$	0.36	\\
GJ 26	&	3484	$\pm$	68	&	M2.5	&	-0.17	$\pm$	0.09	&	0.37	$\pm$	0.07	&	0.37	$\pm$	0.06	&	4.88	$\pm$	0.06	&	-1.741	$\pm$	0.150	&	$<$	0.94			\\
GJ 47	&	3525	$\pm$	68	&	M2	&	-0.26	$\pm$	0.09	&	0.36	$\pm$	0.06	&	0.37	$\pm$	0.06	&	4.88	$\pm$	0.05	&	-1.730	$\pm$	0.135	&	$<$	1.81			\\
GJ 49	&	3712	$\pm$	68	&	M1.5	&	-0.03	$\pm$	0.09	&	0.55	$\pm$	0.05	&	0.53	$\pm$	0.05	&	4.73	$\pm$	0.04	&	-1.317	$\pm$	0.081	&		1.32	$\pm$	0.37	\\
GJ 1030	&	3658	$\pm$	68	&	M2	&	-0.08	$\pm$	0.09	&	0.50	$\pm$	0.05	&	0.49	$\pm$	0.05	&	4.76	$\pm$	0.04	&	-1.409	$\pm$	0.086	&	$<$	0.93			\\
NLTT 4188	&	3810	$\pm$	69	&	M0.5	&	-0.06	$\pm$	0.09	&	0.59	$\pm$	0.06	&	0.57	$\pm$	0.05	&	4.70	$\pm$	0.05	&	-1.213	$\pm$	0.088	&		1.11	$\pm$	0.45	\\
GJ 70	&	3511	$\pm$	68	&	M2.5	&	-0.21	$\pm$	0.09	&	0.37	$\pm$	0.06	&	0.38	$\pm$	0.06	&	4.87	$\pm$	0.06	&	-1.717	$\pm$	0.137	&	$<$	1.02			\\
GJ 3117A	&	3549	$\pm$	68	& M2.5	&	-0.13	$\pm$	0.09	&	0.43	$\pm$	0.06	&	0.43	$\pm$	0.05	&	4.82	$\pm$	0.05	&	-1.588	$\pm$	0.111	&	$<$	0.91			\\
GJ 3126	&	3505	$\pm$	68	&	M3	&	0.01	$\pm$	0.09	&	0.45	$\pm$	0.07	&	0.45	$\pm$	0.06	&	4.80	$\pm$	0.06	&	-1.567	$\pm$	0.118	&	$<$	0.83			\\
GJ 3186	&	3768	$\pm$	68	&	M1	&	-0.14	$\pm$	0.09	&	0.53	$\pm$	0.05	&	0.52	$\pm$	0.05	&	4.74	$\pm$	0.05	&	-1.313	$\pm$	0.088	&	$<$	1.02			\\
GJ 119A	&	3761	$\pm$	69	&	M1	&	-0.08	$\pm$	0.09	&	0.55	$\pm$	0.05	&	0.54	$\pm$	0.05	&	4.72	$\pm$	0.04	&	-1.286	$\pm$	0.085	&	$<$	0.98			\\
GJ119B	&	3508	$\pm$	69	&	M3	&	0.05	$\pm$	0.09	&	0.47	$\pm$	0.06	&	0.46	$\pm$	0.06	&	4.79	$\pm$	0.06	&	-1.535	$\pm$	0.114	&	$<$	0.81			\\
TYC 1795-941-1$^{\ddag}$	&	3774	$\pm$	67	&	M0	&	0.01	$\pm$	0.23	&	0.66	$\pm$	0.13	&	0.64	$\pm$	0.14	&	4.65	$\pm$	0.20	&	-1.128	$\pm$	0.192	&		3.30	$\pm$	0.16	\\
NLTT 10614	&	3728	$\pm$	69	&	M1.5	&	-0.06	$\pm$	0.09	&	0.54	$\pm$	0.05	&	0.53	$\pm$	0.05	&	4.73	$\pm$	0.04	&	-1.315	$\pm$	0.083	&	$<$	2.07			\\
TYC 3720-426-1$^{\ddag}$	&	3822	$\pm$	70	&	M0	&	0.12	$\pm$	0.07	&	0.66	$\pm$	0.1	&	0.64	$\pm$	0.11	&	4.64	$\pm$	0.16	&	-1.106	$\pm$	0.153	&		4.13	$\pm$	0.13	\\
GJ 150.1B	&	3730	$\pm$	68	&	M1	&	-0.16	$\pm$	0.09	&	0.51	$\pm$	0.05	&	0.49	$\pm$	0.05	&	4.76	$\pm$	0.04	&	-1.372	$\pm$	0.088	&		0.87	$\pm$	0.65	\\
GJ 156.1A	&	3745	$\pm$	69	&	M1.5	&	-0.05	$\pm$	0.09	&	0.55	$\pm$	0.05	&	0.54	$\pm$	0.05	&	4.72	$\pm$	0.04	&	-1.289	$\pm$	0.083	&	$<$	2.85			\\
GJ 162	&	3746	$\pm$	68	&	M1	&	-0.19	$\pm$	0.09	&	0.50	$\pm$	0.05	&	0.49	$\pm$	0.05	&	4.77	$\pm$	0.04	&	-1.370	$\pm$	0.090	&		0.93	$\pm$	0.66	\\
GJ 1074	&	3765	$\pm$	69	&	M0.5	&	-0.16	$\pm$	0.09	&	0.52	$\pm$	0.05	&	0.51	$\pm$	0.05	&	4.75	$\pm$	0.05	&	-1.327	$\pm$	0.089	&		1.13	$\pm$	0.50	\\
GJ 184	&	3752	$\pm$	69	&	M0.5	&	-0.10	$\pm$	0.09	&	0.54	$\pm$	0.05	&	0.53	$\pm$	0.05	&	4.73	$\pm$	0.04	&	-1.310	$\pm$	0.086	&	$<$	1.45			\\
GJ 3352	&	3809	$\pm$	69	&	M0.5	&	-0.13	$\pm$	0.09	&	0.56	$\pm$	0.06	&	0.54	$\pm$	0.05	&	4.72	$\pm$	0.05	&	-1.252	$\pm$	0.091	&	$<$	1.47			\\
TYC 3379-1077-1	&	3896	$\pm$	71	&	M0	&	0.04	$\pm$	0.09	&	0.69	$\pm$	0.08	&	0.67	$\pm$	0.07	&	4.61	$\pm$	0.07	&	-1.038	$\pm$	0.099	&		1.85	$\pm$	0.26	\\
TYC7 43-1836-1	&	3846	$\pm$	70	&	M0	&	-0.03	$\pm$	0.09	&	0.62	$\pm$	0.06	&	0.60	$\pm$	0.06	&	4.67	$\pm$	0.06	&	-1.148	$\pm$	0.092	&		1.73	$\pm$	0.30	\\
GJ 272	&	3747	$\pm$	68	&	M1	&	-0.19	$\pm$	0.09	&	0.50	$\pm$	0.05	&	0.49	$\pm$	0.05	&	4.77	$\pm$	0.04	&	-1.368	$\pm$	0.090	&	$<$	1.09			\\
StKM 1-650	&	3874	$\pm$	69	&	M0.5	&	-0.11	$\pm$	0.09	&	0.61	$\pm$	0.07	&	0.60	$\pm$	0.07	&	4.67	$\pm$	0.06	&	-1.144	$\pm$	0.101	&		1.12	$\pm$	0.45	\\
NLTT 21156	&	3616	$\pm$	68	&	M2	&	-0.05	$\pm$	0.09	&	0.50	$\pm$	0.05	&	0.49	$\pm$	0.05	&	4.77	$\pm$	0.04	&	-1.442	$\pm$	0.090	&		0.70	$\pm$	0.61	\\
GJ 399	&	3563	$\pm$	68	&	M2.5	&	0.15	$\pm$	0.10	&	0.55	$\pm$	0.06	&	0.53	$\pm$	0.05	&	4.72	$\pm$	0.05	&	-1.391	$\pm$	0.093	&	$<$	0.88			\\
GJ 408	&	3472	$\pm$	68	&	M2.5	&	-0.19	$\pm$	0.09	&	0.35	$\pm$	0.07	&	0.35	$\pm$	0.06	&	4.89	$\pm$	0.06	&	-1.787	$\pm$	0.163	&	$<$	0.97			\\
GJ 412A	&	3631	$\pm$	68	&	M0.5	&	-0.38	$\pm$	0.09	&	0.38	$\pm$	0.05	&	0.38	$\pm$	0.05	&	4.87	$\pm$	0.04	&	-1.652	$\pm$	0.111	&		1.20	$\pm$	0.82	\\
GJ 414B	&	3661	$\pm$	68	&	M2	&	-0.09	$\pm$	0.09	&	0.50	$\pm$	0.05	&	0.49	$\pm$	0.05	&	4.76	$\pm$	0.04	&	-1.411	$\pm$	0.087	&	$<$	0.94			\\
GJ 3649	&	3691	$\pm$	68	&	M1.5	&	-0.14	$\pm$	0.09	&	0.50	$\pm$	0.05	&	0.49	$\pm$	0.05	&	4.77	$\pm$	0.04	&	-1.406	$\pm$	0.087	&	$<$	1.55			\\
GJ 450	&	3649	$\pm$	68	&	M1.5	&	-0.20	$\pm$	0.09	&	0.45	$\pm$	0.05	&	0.45	$\pm$	0.05	&	4.80	$\pm$	0.04	&	-1.497	$\pm$	0.094	&		1.15	$\pm$	0.51	\\
GJ9404	&	3875	$\pm$	70	&	M0.5	&	-0.10	$\pm$	0.09	&	0.62	$\pm$	0.07	&	0.60	$\pm$	0.07	&	4.67	$\pm$	0.06	&	-1.138	$\pm$	0.101	&		1.25	$\pm$	0.40	\\
GJ 476	&	3498	$\pm$	69	&	M3	&	-0.16	$\pm$	0.09	&	0.38	$\pm$	0.07	&	0.38	$\pm$	0.06	&	4.86	$\pm$	0.06	&	-1.703	$\pm$	0.139	&	$<$	0.93			\\
GJ 9440	&	3710	$\pm$	68	&	M1.5	&	-0.13	$\pm$	0.09	&	0.51	$\pm$	0.05	&	0.50	$\pm$	0.05	&	4.76	$\pm$	0.04	&	-1.378	$\pm$	0.086	&	$<$	0.99			\\
GJ 521A	&	3601	$\pm$	68	&	M1.5	&	-0.09	$\pm$	0.09	&	0.47	$\pm$	0.05	&	0.47	$\pm$	0.05	&	4.79	$\pm$	0.04	&	-1.486	$\pm$	0.094	&	$<$	0.90			\\
GJ 3822	&	3821	$\pm$	70	&	M0.5	&	-0.13	$\pm$	0.09	&	0.56	$\pm$	0.06	&	0.55	$\pm$	0.06	&	4.71	$\pm$	0.05	&	-1.235	$\pm$	0.094	&		0.98	$\pm$	0.55	\\
GJ 548A	&	3903	$\pm$	70	&	M0	&	-0.13	$\pm$	0.09	&	0.63	$\pm$	0.08	&	0.61	$\pm$	0.07	&	4.66	$\pm$	0.07	&	-1.106	$\pm$	0.108	&		1.11	$\pm$	0.47	\\
GJ 552	&	3589	$\pm$	68	&	M2	&	-0.09	$\pm$	0.09	&	0.47	$\pm$	0.05	&	0.46	$\pm$	0.05	&	4.79	$\pm$	0.05	&	-1.503	$\pm$	0.097	&	$<$	0.90			\\
GJ 606	&	3665	$\pm$	68	&	M1.5	&	-0.21	$\pm$	0.09	&	0.46	$\pm$	0.05	&	0.45	$\pm$	0.05	&	4.80	$\pm$	0.04	&	-1.484	$\pm$	0.093	&	$<$	1.57			\\
GJ 3942	&	3867	$\pm$	69	&	M0	&	-0.04	$\pm$	0.09	&	0.63	$\pm$	0.07	&	0.61	$\pm$	0.06	&	4.65	$\pm$	0.06	&	-1.121	$\pm$	0.096	&		1.67	$\pm$	0.30	\\
GJ 625	&	3499	$\pm$	68	&	M2	&	-0.38	$\pm$	0.09	&	0.30	$\pm$	0.07	&	0.31	$\pm$	0.06	&	4.94	$\pm$	0.06	&	-1.894	$\pm$	0.170	&		1.32	$\pm$	0.67	\\
GJ 3997	&	3754	$\pm$	69	&	M0	&	-0.24	$\pm$	0.09	&	0.49	$\pm$	0.05	&	0.48	$\pm$	0.05	&	4.78	$\pm$	0.04	&	-1.391	$\pm$	0.093	&		0.94	$\pm$	0.76	\\
GJ 3998	&	3722	$\pm$	68	&	M1	&	-0.16	$\pm$	0.09	&	0.50	$\pm$	0.05	&	0.49	$\pm$	0.05	&	4.77	$\pm$	0.04	&	-1.382	$\pm$	0.088	&	$<$	1.56			\\
GJ 2128	&	3518	$\pm$	68	&	M2.5	&	-0.30	$\pm$	0.09	&	0.34	$\pm$	0.06	&	0.35	$\pm$	0.06	&	4.90	$\pm$	0.05	&	-1.777	$\pm$	0.144	&	$<$	1.19			\\
GJ 671	&	3422	$\pm$	68	&	M2.5	&	-0.17	$\pm$	0.09	&	0.31	$\pm$	0.09	&	0.32	$\pm$	0.08	&	4.93	$\pm$	0.08	&	-1.909	$\pm$	0.216	&	$<$	0.91			\\
GJ 685	&	3816	$\pm$	69	&	M0.5	&	-0.15	$\pm$	0.09	&	0.55	$\pm$	0.06	&	0.54	$\pm$	0.05	&	4.72	$\pm$	0.05	&	-1.253	$\pm$	0.094	&		1.33	$\pm$	0.42	\\
GJ 686	&	3663	$\pm$	68	&	M1	&	-0.30	$\pm$	0.09	&	0.42	$\pm$	0.05	&	0.42	$\pm$	0.05	&	4.83	$\pm$	0.04	&	-1.548	$\pm$	0.099	&		1.01	$\pm$	0.80	\\
GJ 694.2	&	3847	$\pm$	69	&	M0.5	&	-0.21	$\pm$	0.09	&	0.55	$\pm$	0.06	&	0.54	$\pm$	0.06	&	4.72	$\pm$	0.06	&	-1.241	$\pm$	0.102	&	$<$	1.13			\\
GJ 4057	&	3873	$\pm$	69	&	M0	&	-0.15	$\pm$	0.09	&	0.59	$\pm$	0.07	&	0.58	$\pm$	0.07	&	4.69	$\pm$	0.06	&	-1.167	$\pm$	0.103	&		0.81	$\pm$	0.69	\\
GJ 720A	&	3837	$\pm$	69	&	M0.5	&	-0.14	$\pm$	0.09	&	0.57	$\pm$	0.06	&	0.56	$\pm$	0.06	&	4.71	$\pm$	0.05	&	-1.217	$\pm$	0.096	&	$<$	1.49			\\
GJ 731	&	3844	$\pm$	69	&	M0	&	-0.16	$\pm$	0.09	&	0.57	$\pm$	0.06	&	0.56	$\pm$	0.06	&	4.71	$\pm$	0.06	&	-1.217	$\pm$	0.098	&	$<$	1.59			\\
GJ 740	&	3845	$\pm$	69	&	M0.5	&	-0.14	$\pm$	0.09	&	0.58	$\pm$	0.06	&	0.56	$\pm$	0.06	&	4.70	$\pm$	0.06	&	-1.206	$\pm$	0.097	&		0.92	$\pm$	0.59	\\
GJ 4092	&	3858	$\pm$	69	&	M0.5	&	-0.06	$\pm$	0.09	&	0.62	$\pm$	0.07	&	0.60	$\pm$	0.06	&	4.67	$\pm$	0.06	&	-1.145	$\pm$	0.095	&		1.20	$\pm$	0.41	\\
GJ 9689	&	3824	$\pm$	69	&	M0.5	&	-0.13	$\pm$	0.09	&	0.57	$\pm$	0.06	&	0.55	$\pm$	0.06	&	4.71	$\pm$	0.05	&	-1.231	$\pm$	0.093	&	$<$	1.47			\\
GJ 793	&	3461	$\pm$	68	&	M3	&	-0.21	$\pm$	0.09	&	0.33	$\pm$	0.08	&	0.34	$\pm$	0.07	&	4.91	$\pm$	0.07	&	-1.833	$\pm$	0.176	&	$<$	1.00			\\
BPM 96441	&	3896	$\pm$	72	&	M0	&	-0.03	$\pm$	0.09	&	0.66	$\pm$	0.08	&	0.64	$\pm$	0.07	&	4.63	$\pm$	0.07	&	-1.071	$\pm$	0.103	&		2.05	$\pm$	0.24	\\
TYC 2710-691-1	&	3867	$\pm$	71	&	K7.5		&	0.02	$\pm$	0.09	&	0.65	$\pm$	0.07	&	0.64	$\pm$	0.07	&	4.63	$\pm$	0.06	&	-1.092	$\pm$	0.094	&		2.41	$\pm$	0.21	\\
TYC 2703-706-1	&	3822	$\pm$	70	&	M0.5		&	0.06	$\pm$	0.09	&	0.64	$\pm$	0.06	&	0.62	$\pm$	0.06	&	4.65	$\pm$	0.05	&	-1.136	$\pm$	0.085	&		3.32	$\pm$	0.16	\\
GJ 4196	&	3666	$\pm$	68	&	M1	&	0.07	$\pm$	0.10	&	0.56	$\pm$	0.05	&	0.55	$\pm$	0.05	&	4.71	$\pm$	0.04	&	-1.313	$\pm$	0.082	&		2.40	$\pm$	0.19	\\
NLTT 52021	&	3687	$\pm$	68	&	M2	&	-0.12	$\pm$	0.09	&	0.50	$\pm$	0.05	&	0.49	$\pm$	0.05	&	4.77	$\pm$	0.04	&	-1.400	$\pm$	0.086	&	$<$	0.97			\\
NLTT 53166	&	3832	$\pm$	70	&	M0	&	-0.11	$\pm$	0.09	&	0.58	$\pm$	0.06	&	0.57	$\pm$	0.06	&	4.70	$\pm$	0.05	&	-1.209	$\pm$	0.094	&	$<$	1.45			\\
2MASS J2235$^{\dag}$	&	3891	$\pm$	70	&	K7.5		&	-0.13	$\pm$	0.09	&	0.62	$\pm$	0.07	&	0.60	$\pm$	0.07	&	4.67	$\pm$	0.07	&	-1.127	$\pm$	0.106	&		1.92	$\pm$	0.28	\\
GJ 9793$^{\ddag}$	&	3881	$\pm$	70	&	M0	&	0.24	$\pm$	0.05	&	0.75	$\pm$	0.12	&	0.73	$\pm$	0.12	&	4.58	$\pm$	0.16	&	-0.965	$\pm$	0.146	&		2.77	$\pm$	0.22	\\
GJ 4306	&	3763	$\pm$	69	&	M1	&	-0.13	$\pm$	0.09	&	0.53	$\pm$	0.05	&	0.52	$\pm$	0.05	&	4.74	$\pm$	0.05	&	-1.313	$\pm$	0.088	&	$<$	1.01			\\
GJ 895	&	3748	$\pm$	68	&	M1.5	&	-0.09	$\pm$	0.09	&	0.54	$\pm$	0.05	&	0.53	$\pm$	0.05	&	4.73	$\pm$	0.04	&	-1.308	$\pm$	0.085	&	$<$	1.70			\\
V* BRPsc	&	3553	$\pm$	68	&	M1.5	&	-0.29	$\pm$	0.09	&	0.37	$\pm$	0.06	&	0.37	$\pm$	0.05	&	4.88	$\pm$	0.05	&	-1.704	$\pm$	0.125	&		0.88	$\pm$	0.82	\\
\end{longtable}
\tablefoot{$^{\dag}$ 2MASS J22353504+3712131; $^{\ddag}$ The star falls out of the range of applicability of the metallicity calibrations
 given in \cite{2015A&A...577A.132M}. Metallicities are computed using the photometric calibration by
\cite{2012A&A...538A..25N}, masses from \cite{1993AJ....106..773H}, radius using the calibration by \citet[][Eqn.~4]{2015A&A...577A.132M},
 surface gravities from masses and radius, and luminosities by applying the Stefan-Boltzmann law.
}
}

\section{Rotational Velocities}
\label{rotational_velocities}

  Projected rotational velocities v$\sin i$ have been computed using the cross-correlation
  technique (CCF). Full details on this technique can be found in, e.g.,
  \citet[][]{2001A&A...375..851M,2010A&A...520A..79M}.
  Briefly, for slow rotators, v$\sin i$ $<$ 50 km s$^{\rm -1}$, the CCF can
  be approximated by a Gaussian, and consequently the rotational broadening
  corresponds to a quadratic broadening of the CCF. The observed width
  of the CCF ($\sigma_{\rm obs}$) of a given star when autocorrelated can be written as
  \citep[e.g.][and references therein]{1998A&A...335..183Q}:

  \begin{equation}
  \sigma^{\rm 2}_{\rm obs} = \sigma^{\rm 2}_{\rm rot} + \sigma^{\rm 2}_{\rm 0}
  \end{equation}

  \noindent where $\sigma_{\rm rot}$ is the rotational broadening whilst
  $\sigma_{\rm 0}$ corresponds to the intrinsic CCF width for non-rotating stars.
  $\sigma_{\rm 0}$   includes the intrinsic sources of broadening 
  such as micro and macroturbulence, pressure or Zeeman splitting, and it is dependent on the
  stellar parameters \citep{1998A&A...335..183Q,2010A&A...520A..79M}.
   
  Projected rotational velocity values can be easily obtained from the
  above expression as:

  \begin{equation}
  \label{vsini_eq}
  v\sin i = A\times \sqrt{\sigma^{\rm 2}_{\rm obs} - \sigma^{\rm 2}_{\rm 0}} 
  \end{equation}

   \noindent where A is a constant that depends on the spectrograph.
   To compute A, the spectra of four slowly rotating stars were used,
   namely, GJ 15A, 
   GJ 895, 
   GJ 521, 
   and GJ 552. 
   These stars were selected after checking the available v$\sin i$ values in
   the literature \citep{2010MNRAS.407.1657H}, and have estimates between
   0.52 kms$^{\rm -1}$ (GJ 895) and 1.43 kms$^{\rm -1}$ (GJ 15A).
   We note that since these stars were selected only for the computation
   of the A constant, any star with a low value of v$\sin i$
   can be used.

   The spectra of these stars were broadened up to
   v$\sin i$ = 15 kms$^{\rm -1}$, following the prescriptions provided by
   \cite{2008oasp.book.....G}, using his computation program {\sc SPECTRUM}
   \footnote{http://www.appstate.edu/\textasciitilde{}grayro/spectrum/spectrum.html}.
   A typical value of 0.6 was assumed for the limb-darkening coefficient
   \citep{2008oasp.book.....G,2011A&A...529A..75C}.
   The constant A was found by fitting the relation (v$\sin i$)$^{\rm 2}$ vs.
   $\sigma^{\rm 2}_{\rm obs}$. Only the spectral range 6330-6430 \AA \space
   was used for the CCF. The derived mean value is $<A>$ = 0.476 $\pm$ 0.005.
   We remark that this procedure is commonly used in the literature
   in the cases in which a large sample of stars 
   covering a wide range of (accurate) v$\sin i$ values
   is not available and the targets are expected to be slow rotators (like it is our case).

   In order to model $\sigma_{\rm 0}$ we made use of the
   latest version of the {\sc PHOENIX BT-Settl} atmosphere models
   \citep{Allard2011}.
   A grid of models with T$_{\rm eff}$ between 3000 K and 4000 K was
   computed using the {\sc PHOENIX} web simulator\footnote{http://phoenix.ens-lyon.fr/simulator/}
   assuming $\log g$ = 5.0, and v$\sin i$ equal to zero.  
   It is important to note that the model spectra were synthesised in order to
   match the spectral resolution of the HARPS-N data
   (i.e., $\Delta\lambda$ = 0.01\AA).

   We note that $\log g$ values of 5.0 are adequate for M dwarfs
   \citep[e.g.][]{1996ApJS..104..117L}.
    Three different sets of metallicities were considered.
   Before computing the CCF the synthetic spectra were broadened by convolving it
   with a gaussian profile in order
   to match the instrumental profile of the observed spectra. For this purpose,
   the FWHM of the calibration arc lines were used
   \citep[e.g.][]{2010A&A...520A..79M}.

\begin{figure}
\centering
\includegraphics[angle=270,scale=0.45]{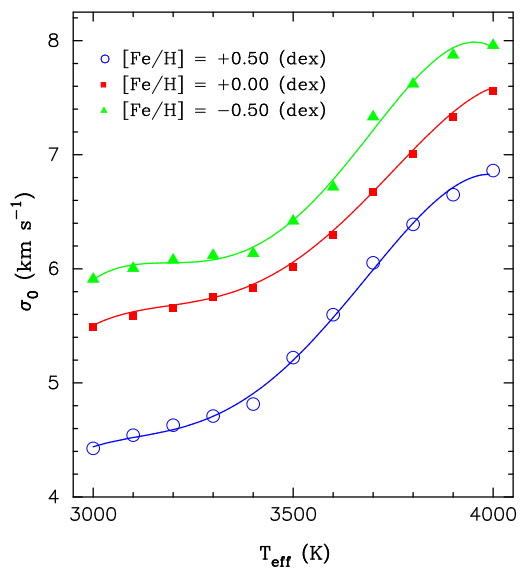}
\caption{
Calibration between the width of the CCF of a non-rotating star, $\sigma_{\rm 0}$, and its effective temperature.
A fourth order polynomial fit is shown.} 
\label{sigma_cero_vs_teff}
\end{figure}

   Each synthetic spectrum was autocorrelated and the width of the CCF,
   $\sigma_{\rm 0}$, was measured by performing a Gaussian fit.
   The dependence of $\sigma_{\rm 0}$ on T$_{\rm eff}$ is shown in
   Figure~\ref{sigma_cero_vs_teff}. The best polynomial fit is shown.
  Once the constant $A$ and $\sigma_{\rm 0}$ for each star are known, rotational velocities
  are derived by measuring $\sigma_{\rm obs}$.

\subsection{Uncertainties and upper limits} 

   Uncertainties in v$\sin i$ were estimated by error propagation. As uncertainty
   in $\sigma_{\rm obs}$, a conservative uncertainty
   of 0.09 km s$^{\rm -1}$ was considered
   as derived from the standard deviation of $\sigma_{\rm obs}$ for stars with more than one observation.
   Regarding the uncertainties in $\sigma_{\rm 0}$, we considered the
   rms of the $\sigma_{\rm 0}$-T$_{\rm eff}$ calibration and
   the errors in the effective temperature (see below).
   We should caution that the errors
   in v$\sin i$ tend to increase towards lower  v$\sin i$ values. 
   Whilst stars with v$\sin i$ larger than 2.0 km s$^{\rm -1}$ show median
   errors of the order of 0.20 km s$^{\rm -1}$, this number increases to
    0.45 km s$^{\rm -1}$ for stars with v$\sin i$ between 1 and 2.0 km s$^{\rm -1}$,
   and  0.65 km s$^{\rm -1}$ for stars with v$\sin i$ below 1 km s$^{\rm -1}$.
    Some stars show large
   errors making their v$\sin i$ values compatible with zero. For these stars
   we provided upper limits (computed as v$\sin i$ + $\Delta$v$\sin i$).

   We are aware that a more detailed error analysis need a comprehensive study of the
   dependence of $\sigma_{\rm obs}$ on the signal-to-noise ratio (S/N)
   as well as on the depth of the CCF \citep[see e.g.][and references therein]{2001A&A...375..851M}.
   In particular, our trend of lower uncertainties towards higher v$\sin i$ may be influenced
   by the fact that higher v$\sin i$ values translate into lower CCF depths and, therefore, higher
   errors on $\sigma_{\rm obs}$.
   Since the dependence of  $\sigma_{\rm obs}$ on parameters such as
   the spectra S/N or the CCF depth is also crucial to determine
   the errors when measuring radial velocities (one of the main purposes
   of the HADES survey), we let such a study for a forthcoming work. 
   In any case we remark that our estimated uncertainties are compatible with the uncertainties
   reported in the literature when using the CCF technique, typically
   in the range 0.3-0.6 km s$^{\rm -1}$ \citep[e.g.][]{2010AJ....139..504B}.

   Since we are considering stars with very slow rotation our capability of measuring
   the projected rotational velocity is linked to an accurate determination of
   $\sigma_{\rm 0}$. It might be the case that for very slow rotators
   ($\sigma_{\rm obs}$ $\backsimeq$ $\sigma_{\rm 0}$)
   our calibration returns a $\sigma_{\rm 0}$ value slightly larger than $\sigma_{\rm obs}$.
   In these cases upper limits were determined as follows.
   For slow rotators we can write
   $\sigma_{\rm obs}$ $=$ $\sigma_{\rm 0}$ + $\epsilon$  where
   $\epsilon$ $\ll$  $\sigma_{\rm 0}$, and therefore
   Eq.~\ref{vsini_eq} leads to:

   \begin{equation}
   \label{vsini_upp}
   v\sin i  \le  A\times\sqrt{2\sigma_{\rm 0}\epsilon} 
   \end{equation}

   \noindent As we are in the very slow rotation domain,
   it is reasonable to assume $\epsilon$ $\sim$
   $\Delta\sigma_{\rm 0}$. Two main sources of uncertainty in
   $\sigma_{\rm 0}$ were considered: {\it i)} the errors associated to the stellar effective temperature,
   which are of the order of 70 K; and {\it ii)} the errors associated to the $\sigma_{\rm 0}$-T$_{\rm eff}$
   calibration for which we consider its corresponding rms. 
   The derived $v\sin i$ values, errors, and upper limits are listed in Table~\ref{parameters_table_full}.

\section{Spectral subtraction}
\label{spectral_subtraction}
\subsection{Reference ``inactive'' stars}

 In order to determine the emission excess in the different
 chromospheric indicators we subtract
 the underlying photospheric contribution from the stellar spectrum.
 To do this we employed the spectral subtraction technique
 \citep[e.g.][]{1994A&A...284..883F,1995A&AS..114..287M,2000A&AS..146..103M}.
 This technique automatically subtracts the basal
 chromospheric flux provided that the spectrum of a non-active
 star of similar stellar parameters and chemical composition
 to the target star is used as reference
 \citep{2010A&A...520A..79M}.

 To select our quiet templates for each star and
 each observation the Ca~{\sc ii} H \& K  S index was computed. 
 Our definition of the bandpasses for the S index is made
 following \cite{1996AJ....111..439H}.
 The fluxes in the central cores of the  Ca~{\sc ii} H \& K lines
 are measured in two 3.28 \AA \space wide windows centred at
 3968.47 \AA \space and 3933.67 \AA \space  respectively. 
 Continuum fluxes on the sides of the lines are measured in two 20 \AA
 \space windows with central wavelengths at
 3901.07 and 4001.07 \AA.  
 Fluxes were measured using the IRAF\footnote{
 IRAF is distributed by the National Optical Astronomy Observatories,
 which are operated by the Association of Universities for Research
 in Astronomy, Inc., under cooperative agreement with the National
 Science Foundation.} task {\sc sbands}. Before measuring the fluxes,
 each individual spectrum was corrected for its corresponding radial
 velocity using the IRAF task {\sc dopcor}. No attempt to convert
 our S index into the Mount Wilson scale or to correct it from
 the underlying stellar photospheric contribution was done.
 We note that although \cite{2015MNRAS.452.2745S}  extended the original R$^{'}_{\rm HK}$ calibration
 by \cite{1984ApJ...279..763N}
 up to $(B-V)$ $\sim$ 1.9, the use of R$^{'}_{\rm HK}$ is not 
 needed for the purpose of this work. 

\begin{figure}[!htb]
\centering
\includegraphics[angle=270,scale=0.45]{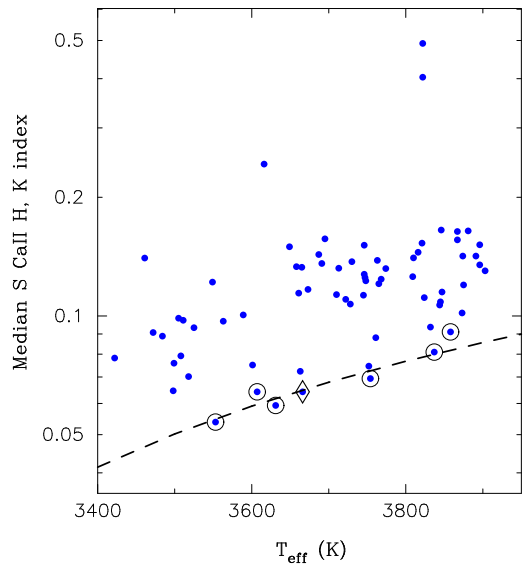}
\caption{Median S index for each star vs. stellar effective temperature.
 The dashed line represents a fit to those stars with the lowest S values
 (shown with circles). The star GJ 4196, discussed in the text, is shown with
 a diamond symbol.
 }
\label{medianS_vs_teff}
\end{figure}

 Figure~\ref{medianS_vs_teff} shows for each star the median S index value 
 as a function of the stellar effective temperature. 
 Given that our sample covers a wide range of S index values, it is
 reasonable to assume that the stars with the lowest S index
 (those stars lying on the dashed line in figure~\ref{medianS_vs_teff}) are the ``least active'' 
 stars in our sample. These stars
 (namely GJ 15A, GJ 184, GJ 412A, GJ 720A, GJ 3997, GJ 4092, and V* BR Psc)
 were selected 
 as references for the spectral subtraction for all the activity indicators.

 The star GJ 4196 also lies among the lowest S index stars in the sample
 but was not selected as reference
 for the spectral subtraction
 given that it has a metallicity value significantly larger than the remaining
 reference stars, see next subsection.
 It is worth noticing that metallicity effects are usually not taking into account in the
 computation of the S index, although \citet{2011arXiv1107.5325L} noticed that 
 for metal-poor stars the continuum passbands are weaker, resulting in slightly
 larger S values. 
 

  
\subsection{Emission excess fluxes}

 An extensive description of the procedure adopted to compute the excess fluxes is
 provided by \citet[][submitted]{Gaetano}. 
 Here we give a summary of the
 reduction steps that lead to the measurement of the flux excess.

 The spectra provided by the  DRS show night-to-night
 variations in the continuum level at different wavelengths, due to atmospheric
 differential absorption and instrumental effects. To correct them, and to scale
 the observed spectra to the same flux reference, we compare them with
 synthetic spectra from the BT-Settl spectral library provided by \citet{Allard2011}.
 The library was interpolated in order to obtain for each star a model atmosphere
 with its corresponding stellar parameters (T$_{\rm eff}$, $\log g$, and $[Fe/H]$).
 Both the observed spectra and the model were degraded to a low-resolution
 (down to R$<$50) in order to avoid discrepancies between the observed and
 the model lines profiles. Finally, the spectrum-to-model flux ratio was used to
 rescale the observed high-resolution spectrum.
 The flux-rescaled spectra are then corrected for telluric contamination using
 the spectrum of the telluric standard $\eta$ UMa 
 observed with the same spectrograph used in this work by our group
 within the context of the GAPS program \citep[see][]{2015A&A...578A..64B}.
 For each star, we also compute the median of all the corresponding  observed spectra.

 For each star in our sample, we interpolate the grid of the median spectra of
 the selected reference stars (see previous section) over T$_{\rm eff}$.
 Before performing the interpolations, the rotational broadening (as measured
 in this work) of both the observed and reference spectra are applied. 
 The rotational broadening is performed by convolving each stellar spectrum with the required rotational
  profile, as provided by
  \cite{2008oasp.book.....G}. 
 Stellar metallicity is also taken into account as 
 our reference stars have relatively low metallicities. 
 This is in part because of the fact that the continuum flux predicted by the
 {\sc BT-Settl} atmosphere models decreases towards lower metallicities.
 The result is an offset between the observed and template fluxes with the observed
 fluxes being lower than the references ones.
 In order to correct
 for this offset, a low-order polynomial fit is applied.
 Finally, line flux excesses were computed by integrating the difference spectrum in
 the wavelength ranges listed in Table~\ref{tab:indicatorsJesus}.
 These ranges were set after a visual inspection of the subtracted spectra.
 Errors in the flux excesses were estimated by propagating the S/N of the subtracted
 spectrum out of the core of the lines.

\begin{table}
\begin{center}
\caption{Chromospheric emission lines analysed in this work.
 The widths of the spectral windows 
 to compute the emission fluxes were set after a visual inspection of the spectra.
}\label{tab:indicatorsJesus}
\begin{tabular}{lccc}
\hline\hline
Ion & Line & Central wavelength (\AA) & Spectral width (\AA)\\
\hline
Ca~{\sc ii} & K & 3933.67 & 1.5 \\
Ca~{\sc ii} & H & 3968.47 & 1.5 \\
\hline
H~{\sc i} & H$\epsilon$ & 3970.07 & 1.5 \\
H~{\sc i} & H$\delta$   & 4101.76 & 1.5 \\
H~{\sc i} & H$\gamma$   & 4340.46 & 1.5 \\
H~{\sc i} & H$\beta$    & 4861.32 & 3.0 \\
H~{\sc i} & H$\alpha$   & 6562.80 & 4.0 \\
\hline
\end{tabular}
\end{center}
\end{table}

  It is worth noticing that for some stars showing calcium emission we were not
  able to measure any H${\alpha}$ emission.
  This might be related to the complex mechanisms involved in the H${\alpha}$ emission.
  Note that we are considering emission excesses in the subtracted
  spectrum, and it was shown that  at relatively
  low-activity levels (such as the ones of our sample) the flux radiated in the
  H$\alpha$ line seems to initially decrease  with increasing calcium flux, resulting
  in an H$\alpha$ extra-absorption
  \citep[][submitted]{1990ApJS...74..891R,2009AJ....137.3297W,Gaetano}.

\section{Other stellar properties}
\label{spectral_others}
\subsection{Spatial velocity components and age}
\label{spatial_velocity_subsec}

  Stellar age is one of the most difficult parameters
  to obtain accurately. A ``rough'' 
  age estimate can be obtained if the star is a member
  of a stellar kinematic group or a young stellar association.
  Indeed, it seems that most of the active early-M dwarfs may belong to
  young associations \citep{2012AJ....143...93R}.
  However, it should be noted that identifying stars in kinematic groups
  is not a trivial task. Lists of members change among different works
  and many old stars can share the spatial motion of young stars in
  kinematic groups.
   For example, \cite{2009A&A...499..129L}
  show that among previous lists of Local Association members,
  roughly 30\% are old field stars.
  Therefore, kinematic criteria alone are not sufficient 
  to conclude about the young nature of a star on a robust basis.
  Usually a combination of kinematics, spectroscopic signatures
  of youth (e.g. rotation, activity), and the location of the
  stars in colour-magnitude diagrams are used to assess the
  likelihood of membership of a star to young kinematic groups
  \citep[e.g.][]{2001A&A...379..976M,2010A&A...521A..12M}.

  Galactic spatial-velocity components $(U, V, W)$ were computed for our targets
  using the mean radial velocity measured within the HADES project
  together with parallaxes and proper motions
  \citep{1995AJ....110.1838R,2005AJ....130.1680L}.
  To compute $(U, V, W)$ we followed the procedure of \cite{2001MNRAS.328...45M}
  and \cite{2010A&A...521A..12M}.
  It is known that binarity might alter the derived kinematic properties
  in the case of close-in binaries.
  It is, however, more unlikely that for wide visual binaries the classification 
  of a system as kinematically old/young might be affected. Further, 
  catalogues of binaries might include many optical (non-physical) systems. 
  We, therefore, kept in the analysis
  the stars in binary systems listed in the CDC \citep{2002yCat.1274....0D},
  the WVDSC catalogue \citep{2001AJ....122.3466M}, or in Simbad \citep{2000A&AS..143....9W}.
  A total of 16 stars ($\sim$ 23\% of the sample) have an entry in at least one of these catalogues.
  In order to identify close-in spectroscopic binaries,
  the SB9 \citep{2004A&A...424..727P} and CAB3 \citep{2008MNRAS.389.1722E}
  catalogues were searched, 
  but no match was found.  
   We note that as the HADES project is an exoplanet survey, 
  those stars for which signatures of a close-in stellar companion have been found 
  were excluded from the survey and not considered for further follow up.

\begin{figure*}[!htb]
\centering
\includegraphics[angle=270,scale=0.525]{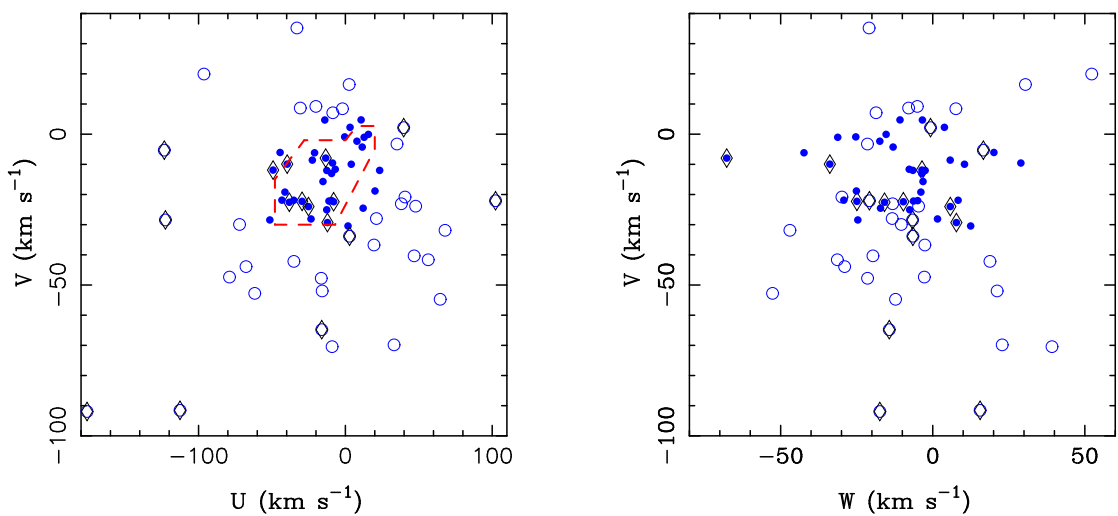}
\caption{
 $(U, V)$ and $(W, V)$ planes for the observed stars. 
 The dashed red line represents the boundary of the young disc population as defined by
\cite{1984AJ.....89.1358E,1989PASP..101..366E}. Stars inside or close to this
 boundary are shown with filled circles. Stars flagged as binaries are shown with
diamonds.} 
\label{uvw_planes}
\end{figure*}

  Figure~\ref{uvw_planes} shows the $(U, V)$ and $(W, V)$ planes.
  We identified as  kinematically ``young'' those stars inside or near the
  boundary of the young disc population as defined by
  \cite{1984AJ.....89.1358E,1989PASP..101..366E}.
  A total of 37 stars (roughly 51\% of the whole sample) were classified as
   kinematically ``young'' (possible ages
  $\lesssim$ 650 Myr, i.e., age of the Hyades open cluster).
  The fraction of possible young stars among the single stars is 29/55 (i.e., $\sim$ 53\%)
  while among the binaries, 8/16 (50\%) of the stars are kinematically young. 
  Our derived $(U, V, W)$ velocities are given in Table~\ref{kinematic_catalogue}.

\onllongtab{
\begin{longtab}
\begin{landscape}
\begin{longtable}{lcccccccccr}
\caption{
{
 Kinematic data for the observed stars. }
  }\label{kinematic_catalogue}\\
\hline
\hline
 Star & $\alpha$ & $\delta$       & $\mu_{\alpha}$         & $\mu_{\delta}$         & $\pi^{\dag}$ &  RV              &  $U$ & $V$ & $W$ & Notes$^{\ddag}$ \\
      & (h,m,s)  & ($^{o}$,','')  & (arcsec yr$^{\rm -1}$) &   (arcsec yr$^{\rm -1}$) & (arcsec) & kms$^{\rm -1}$ & kms$^{\rm -1}$ & kms$^{\rm -1}$ & kms$^{\rm -1}$ & \\
 (1) & (2)  &  (3) & (4) & (5) & (6) & (7) & (8) & (9) & (10) & (11)  \\            
\hline
\endfirsthead
\caption{Continued.} \\
\hline
 Star & $\alpha$ & $\delta$       & $\mu_{\alpha}$         & $\mu_{\delta}$         & $\pi^{\dag}$ &  RV              &  $U$ & $V$ & $W$ & Notes$^{\ddag}$ \\
      & (h,m,s)  & ($^{o}$,','')  & (arcsec yr$^{\rm -1}$) &   (arcsec yr$^{\rm -1}$) & (arcsec) & kms$^{\rm -1}$ & kms$^{\rm -1}$ & kms$^{\rm -1}$ & kms$^{\rm -1}$ & \\
 (1) & (2)  &  (3) & (4) & (5) & (6) & (7) & (8) & (9) & (10) & (11)  \\
\hline
\endhead
\hline
\endfoot
\hline
\endlastfoot
GJ 2	&	00:05:10.2	&	+45:47:13	&	0.870	&	-0.151	&	0.0889	$\pm$	0.0014	a	&	-0.04	&	-38.15	&	-22.58	&	-15.86	&	CCDM/WVDSC YD	\\
GJ 3014	&	00:13:37.8	&	+80:39:55	&	0.251	&	0.182	&	0.0510	$\pm$	0.0018	a	&	-15.39	&	-12.20	&	-29.29	&	7.77	&	WVDSC YD	\\
GJ 16	&	00:18:16.6	&	+10:12:10	&	0.000	&	-0.020	&	0.0463	$\pm$	0.0069	c	&	-14.85	&	4.02	&	-9.98	&	10.43	&	YD	\\
GJ 15A	&	00:18:20.5	&	+44:01:19	&	2.888	&	0.409	&	0.2803	$\pm$	0.0010	e	&	12.00	&	-49.23	&	-11.90	&	-3.54	&	CCDM/WVDSC YD	\\
GJ 21	&	00:26:52.9	&	+70:08:34	&	-0.135	&	-0.148	&	0.0607	$\pm$	0.0014	a	&	-2.78	&	10.69	&	4.74	&	-10.76	&	YD	\\
GJ 26	&	00:38:58.0	&	+30:36:57	&	1.556	&	0.032	&	0.0801	$\pm$	0.0160	b	&	-0.17	&	-78.94	&	-47.38	&	-2.76	&		\\
GJ 47	&	01:01:19.6	&	+61:22:02	&	0.368	&	-0.823	&	0.0909	$\pm$	0.0182	b	&	7.76	&	-20.98	&	-6.16	&	-42.34	&	YD	\\
GJ 49	&	01:02:38.0	&	+62:20:41	&	0.730	&	0.089	&	0.1004	$\pm$	0.0015	a	&	-5.78	&	-25.10	&	-24.00	&	5.75	&	WVDSC YD	\\
GJ 1030	&	01:06:41.6	&	+15:16:24	&	-0.112	&	-0.254	&	0.0452	$\pm$	0.0023	a	&	17.12	&	12.76	&	-1.02	&	-31.25	&	YD	\\
NLTT 4188	&	01:16:10.1	&	+60:09:13	&	0.376	&	-0.201	&	0.0320	$\pm$	0.0090	f	&	8.30	&	-51.40	&	-28.45	&	-24.64	&	YD	\\
GJ 70	&	01:43:20.4	&	+04:19:24	&	-0.420	&	-0.763	&	0.0876	$\pm$	0.0020	a	&	-25.71	&	47.87	&	-23.86	&	-4.66	&		\\
GJ 3117A	&	01:51:50.8	&	+64:26:07	&	0.247	&	-0.190	&	0.0660	$\pm$	0.0260	f	&	-12.60	&	-8.03	&	-22.41	&	-9.66	&	CCDM/WVDSC YD	\\
GJ 3126	&	02:01:35.3	&	+63:46:12	&	-0.255	&	-0.090	&	0.0784	$\pm$	0.0157	b	&	-83.94	&	64.54	&	-54.75	&	-12.22	&		\\
GJ 3186	&	02:52:25.0	&	+26:58:32	&	0.010	&	-0.232	&	0.0362	$\pm$	0.0024	a	&	-10.86	&	12.09	&	-24.54	&	-17.16	&	YD	\\
GJ 119A	&	02:56:34.4	&	+55:26:14	&	0.733	&	-0.452	&	0.0222	$\pm$	0.0011	a	&	76.84	&	-175.96	&	-91.95	&	-17.39	&	CCDM/WVDSC	\\
GJ 119B	&	02:56:35.2	&	+55:26:30	&	0.803	&	-0.444	&	0.0433	$\pm$	0.0087	b	&	76.03	&	-122.55	&	-28.42	&	-6.64	&	CCDM	\\
TYC 1795-941-1	&	03:12:12.6	&	+29:51:32	&	-0.041	&	-0.074	&	0.0241	$\pm$	0.0054	d	&	-21.33	&	23.34	&	-11.99	&	-6.55	&	YD	\\
NLTT 10614	&	03:20:45.2	&	+39:43:01	&	0.125	&	-0.096	&	0.0450	$\pm$	0.0120	f	&	5.88	&	-12.62	&	-12.03	&	-2.50	&	YD	\\
TYC 3720-426-1	&	03:41:37.3	&	+55:13:07	&	0.095	&	-0.119	&	0.0284	$\pm$	0.0022	a	&	-3.29	&	-11.19	&	-22.17	&	-6.32	&	YD	\\
GJ 150.1B	&	03:43:45.2	&	+16:40:02	&	0.159	&	-0.313	&	0.0599	$\pm$	0.0024	a	&	34.86	&	-29.49	&	-22.26	&	-24.94	&	CCDM YD	\\
GJ 156.1A	&	03:56:47.4	&	+53:33:37	&	0.309	&	-0.406	&	0.0371	$\pm$	0.0033	a	&	-20.33	&	-16.06	&	-64.81	&	-14.31	&	CCDM/WVDSC	\\
GJ 162	&	04:08:37.4	&	+33:38:13	&	0.525	&	0.126	&	0.0744	$\pm$	0.0027	a	&	35.14	&	-44.48	&	-6.06	&	20.09	&	YD	\\
GJ 1074	&	04:58:46.0	&	+50:56:38	&	0.503	&	-0.336	&	0.0518	$\pm$	0.0023	a	&	17.30	&	-35.10	&	-42.17	&	18.80	&		\\
GJ 184	&	05:03:23.8	&	+53:07:43	&	1.304	&	-1.537	&	0.0734	$\pm$	0.0020	a	&	65.95	&	-112.58	&	-91.54	&	15.58	&	WVDSC	\\
GJ 3352	&	05:34:08.7	&	+51:12:56	&	-0.052	&	-0.222	&	0.0392	$\pm$	0.0026	a	&	-71.68	&	56.40	&	-41.64	&	-31.37	&		\\
TYC 3379-1077-1	&	06:14:42.4	&	+47:27:35	&	-0.082	&	-0.031	&	0.0270	$\pm$	0.0070	f	&	29.04	&	-30.69	&	8.68	&	-7.93	&		\\
TYC 743-1836-1	&	06:19:29.5	&	+13:57:03	&	0.105	&	-0.044	&	0.0390	$\pm$	0.0110	f	&	39.89	&	-35.13	&	-21.90	&	8.32	&	YD	\\
GJ 272	&	07:23:14.9	&	+46:05:15	&	-0.117	&	-0.240	&	0.0609	$\pm$	0.0019	a	&	-30.99	&	20.18	&	-18.84	&	-25.10	&	YD	\\
StKM 1-650	&	07:31:36.1	&	+62:01:11	&	0.026	&	-0.153	&	0.0250	$\pm$	0.0070	f	&	-18.15	&	2.87	&	-33.83	&	-6.54	&	WVDSC	\\
NLTT 21156	&	09:13:23.8	&	+68:52:31	&	-0.156	&	-0.231	&	0.0650	$\pm$	0.0180	f	&	14.08	&	-22.47	&	-8.59	&	5.72	&	YD	\\
GJ 399	&	10:39:41.0	&	-06:55:24	&	-0.716	&	-0.108	&	0.0608	$\pm$	0.0031	a	&	3.43	&	-43.20	&	-21.88	&	-29.23	&	YD	\\
GJ 408	&	11:00:04.5	&	+22:50:01	&	-0.426	&	-0.280	&	0.1501	$\pm$	0.0017	a	&	3.34	&	-9.35	&	-13.04	&	-3.59	&	YD	\\
GJ 412A	&	11:05:28.6	&	+43:31:36	&	-4.411	&	0.943	&	0.2063	$\pm$	0.0010	a	&	69.09	&	-123.34	&	-5.31	&	16.64	&	Double-Mul	\\
GJ 414B	&	11:11:01.9	&	+30:26:44	&	0.609	&	-0.208	&	0.0830	$\pm$	0.0166	b	&	-15.16	&	39.69	&	2.16	&	-0.72	&	Double-Mul	\\
GJ 3649	&	11:12:38.9	&	+18:56:04	&	0.000	&	0.000	&	0.0694	$\pm$	0.0201	c	&	31.74	&	-8.75	&	-9.57	&	28.98	&	YD	\\
GJ 450	&	11:51:07.5	&	+35:16:17	&	-0.272	&	0.255	&	0.1165	$\pm$	0.0012	a	&	0.47	&	-14.03	&	4.69	&	-3.39	&	YD	\\
GJ 9404	&	12:19:24.5	&	+28:22:55	&	-0.650	&	0.076	&	0.0394	$\pm$	0.0019	a	&	-0.50	&	-72.09	&	-29.92	&	-10.35	&		\\
GJ 476	&	12:35:01.0	&	+09:49:45	&	-0.449	&	-0.319	&	0.0547	$\pm$	0.0030	a	&	33.47	&	-15.80	&	-51.97	&	21.17	&		\\
GJ 9440	&	13:19:40.3	&	+33:20:48	&	-0.299	&	-0.144	&	0.0589	$\pm$	0.0015	a	&	-11.56	&	-12.70	&	-25.07	&	-7.58	&	YD	\\
GJ 521A	&	13:39:24.1	&	+46:11:08	&	-0.043	&	0.391	&	0.0769	$\pm$	0.0016	a	&	-65.16	&	-13.42	&	-7.94	&	-67.75	&	Double-Mul YD	\\
GJ 3822	&	14:02:19.6	&	+13:41:23	&	0.097	&	-0.140	&	0.0504	$\pm$	0.0020	a	&	-7.89	&	11.46	&	-4.26	&	-13.02	&	YD	\\
GJ 548A	&	14:25:43.5	&	+23:37:02	&	0.794	&	-1.116	&	0.0611	$\pm$	0.0014	a	&	9.63	&	102.29	&	-22.06	&	-20.85	&	Double-Mul	\\
GJ 552	&	14:29:30.3	&	+15:31:46	&	-1.053	&	1.300	&	0.0714	$\pm$	0.0021	a	&	8.04	&	-96.29	&	19.93	&	52.27	&		\\
GJ 606	&	15:59:53.3	&	-08:15:11	&	0.204	&	-0.023	&	0.0720	$\pm$	0.0019	a	&	-16.95	&	-8.53	&	7.09	&	-18.62	&		\\
GJ 3942	&	16:09:02.9	&	+52:56:37	&	0.204	&	0.061	&	0.0591	$\pm$	0.0010	a	&	-18.71	&	-0.47	&	-0.89	&	-25.31	&	YD	\\
GJ 625	&	16:25:24.2	&	+54:18:16	&	0.432	&	-0.171	&	0.1535	$\pm$	0.0010	a	&	-12.85	&	7.90	&	-2.35	&	-17.41	&	YD	\\
GJ 3997	&	17:15:50.2	&	+19:00:00	&	-0.130	&	0.015	&	0.0763	$\pm$	0.0162	d	&	-20.57	&	-15.24	&	-15.70	&	-3.20	&	YD	\\
GJ 3998	&	17:16:00.7	&	+11:03:30	&	-0.138	&	-0.349	&	0.0562	$\pm$	0.0023	a	&	-44.81	&	-16.44	&	-47.76	&	-21.42	&		\\
GJ 2128	&	17:16:41.2	&	+08:03:30	&	-0.281	&	-0.067	&	0.0671	$\pm$	0.0027	a	&	-30.40	&	-23.41	&	-28.10	&	1.57	&	YD	\\
GJ 671	&	17:19:52.5	&	+41:42:57	&	0.285	&	-0.822	&	0.0808	$\pm$	0.0017	a	&	-19.32	&	40.64	&	-20.84	&	-29.87	&		\\
GJ 685	&	17:35:34.2	&	+61:40:58	&	0.264	&	-0.514	&	0.0709	$\pm$	0.0010	a	&	-14.70	&	35.18	&	-3.26	&	-21.46	&		\\
GJ 686	&	17:37:52.8	&	+18:35:21	&	0.927	&	0.983	&	0.1237	$\pm$	0.0016	a	&	-9.33	&	-33.02	&	35.20	&	-20.93	&		\\
GJ 694.2	&	17:45:33.6	&	+46:51:19	&	-0.021	&	-0.021	&	0.0476	$\pm$	0.0038	c	&	4.61	&	3.16	&	2.28	&	3.84	&	YD	\\
GJ 4057	&	18:25:04.8	&	+24:38:08	&	-0.041	&	-0.447	&	0.0458	$\pm$	0.0019	a	&	0.86	&	38.11	&	-23.07	&	-13.21	&		\\
GJ 720A	&	18:35:18.0	&	+45:44:35	&	0.452	&	0.365	&	0.0643	$\pm$	0.0010	a	&	-31.31	&	-39.66	&	-9.92	&	-33.82	&	Double-Mul YD	\\
GJ 731	&	18:51:51.3	&	+16:35:04	&	-0.226	&	-0.483	&	0.0646	$\pm$	0.0016	a	&	-14.34	&	19.51	&	-36.74	&	-2.59	&		\\
GJ 740	&	18:58:00.2	&	+05:54:39	&	-0.194	&	-1.222	&	0.0917	$\pm$	0.0015	a	&	10.61	&	46.79	&	-40.34	&	-19.67	&		\\
GJ 4092	&	18:59:38.4	&	+07:59:15	&	0.364	&	-0.181	&	0.0385	$\pm$	0.0023	a	&	-82.80	&	-61.69	&	-52.76	&	-52.66	&		\\
GJ 9689	&	20:13:51.8	&	+13:23:20	&	0.423	&	0.020	&	0.0382	$\pm$	0.0027	a	&	-67.70	&	-67.67	&	-43.90	&	-28.97	&		\\
GJ 793	&	20:30:31.4	&	+65:26:55	&	0.443	&	0.284	&	0.1251	$\pm$	0.0011	a	&	10.78	&	-20.09	&	9.20	&	-5.07	&		\\
BPM 96441	&	21:12:55.4	&	+31:07:54	&	0.002	&	-0.115	&	0.0260	$\pm$	0.0070	f	&	6.42	&	15.66	&	-0.06	&	-15.35	&	YD	\\
TYC 2710-691-1	&	21:17:59.1	&	+34:04:30	&	0.058	&	-0.018	&	0.0260	$\pm$	0.0070	f	&	-10.99	&	-6.92	&	-11.65	&	-7.74	&	YD	\\
TYC 2703-706-1	&	21:18:33.7	&	+30:14:35	&	0.057	&	-0.021	&	0.0259	$\pm$	0.0061	d	&	-21.71	&	-9.23	&	-22.02	&	-4.97	&	YD	\\
GJ 4196	&	21:27:33.0	&	+34:01:29	&	-0.277	&	-0.182	&	0.0359	$\pm$	0.0031	a	&	-67.70	&	33.18	&	-69.83	&	22.84	&		\\
NLTT 52021	&	21:44:54.0	&	+44:17:09	&	-0.143	&	-0.655	&	0.0378	$\pm$	0.0025	a	&	-27.59	&	67.88	&	-31.87	&	-46.97	&		\\
NLTT 53166	&	22:11:17.0	&	+41:00:55	&	-0.207	&	0.231	&	0.0440	$\pm$	0.0022	a	&	9.36	&	2.58	&	16.45	&	30.45	&		\\
2MASS J22353504+3712131	&	22:35:35.0	&	+37:12:13	&	-0.020	&	0.044	&	0.0230	$\pm$	0.0060	f	&	5.76	&	-1.98	&	8.42	&	7.59	&		\\
GJ 9793 	&	22:41:35.3	&	+18:49:28	&	0.240	&	0.056	&	0.0275	$\pm$	0.0022	a	&	-16.58	&	-41.17	&	-19.21	&	-3.91	&	YD	\\
GJ 4306	&	22:55:59.9	&	+17:48:40	&	0.025	&	-0.109	&	0.0600	$\pm$	0.0160	f	&	-31.72	&	1.79	&	-30.42	&	12.47	&	YD	\\
GJ 895	&	23:24:30.6	&	+57:51:18	&	-0.063	&	-0.283	&	0.0772	$\pm$	0.0013	a	&	-33.00	&	21.14	&	-27.96	&	-13.33	&		\\
V* BR Psc	&	23:49:11.9	&	+02:24:12	&	0.995	&	-0.968	&	0.1673	$\pm$	0.0012	a	&	-70.95	&	-9.02	&	-70.44	&	39.21	&		\\
\end{longtable}
\tablefoot{$^{\dag}$ a) \citet{2007A&A...474..653V}; b) NStars database; c) \citet{1997AJ....113.1458H};
 d) \citet{2016yCat.1333....0F};
 e) \citet{1997ESASP1200.....E};
 f) Spectroscopic parallax from \citet{2013AJ....145..102L}\\
$^{\ddag}$ WVDSC: The Washington Visual Double Star Catalog \citep{2001AJ....122.3466M}; CCDM: Catalog of Components of Double \& Multiple stars  \citep{2002yCat.1274....0D}; YD: Possible Young Disc star; Rest of notes are from the Simbad database.}
\end{landscape}
\end{longtab}
}

\subsection{X-ray fluxes} 

 We searched for X-ray counterparts by collecting the count rates
 and hardness ratio data provided by the {\sc HEASARC}\footnote{http://heasarc.nasa.gov/docs/archive.html} archive
 determined from the PSPC instrument on board of the {\it ROSAT}
 mission \citep{1999A&A...349..389V,2000IAUC.7432R...1V}.
 To determine the X-ray fluxes
 we used the count rate-to-energy flux conversion factor ($C_{\rm X}$) relation given
 by \cite{1995ApJ...450..392S}:

 \begin{equation}
 C_{\rm X} = (8.31 + 5.30 \ \textrm{HR})  10^{-12} \textrm{erg} \ \textrm{cm}^{-2} \ \textrm{counts}^{-1}.
 \end{equation}

 \noindent where HR is the hardness ratio of the star in the ROSAT energy band 0.1-2.4 KeV,
 defined as HR = $(H-S)/(H+S)$ where $H$ and $S$ refers to counts in the hard (0.5-2.0 KeV),
 and soft (0.1-0.4 KeV) bands, respectively. 
 Combining the X-ray count rate, $f_{\rm X}$ $(\textrm{counts \ s}^{-1})$, and the conversion factor $C_{\rm X}$ with
 the distance $d$ (pc), the stellar X-ray luminosity $L_{\rm X}$ $(\textrm{erg\ s}^{-1})$ can be estimated.
 This approach assumes that absorption effects are not of significant importance as our targets
 are nearby (d $\lesssim$ 45 pc).
 For two targets (GJ 9440 and GJ 476) X-ray fluxes were taken directly from the XMM 
 XAssist Source List \citep{2003ASPC..295..465P}.

\section{Results}
\label{results}

  Our stellar sample is presented in Table~\ref{parameters_table_full} where the basic stellar
  parameters are listed. 
  Kinematic and ancillary data are shown in Table~\ref{kinematic_catalogue}.
  Finally, our derived emission excesses are listed in Table~\ref{parameters_emission_excesses}. 
  These three tables are available online. 


\onllongtab{
\begin{longtab}
\begin{landscape}
\begin{longtable}{lcccccccc}
\caption{
{
Emission excesses, $\log$ F$_{\lambda}$ [erg cm$^{\rm -2}$ s$^{\rm -1}$] }
  }\label{parameters_emission_excesses}\\
\hline
\hline
 Star & $\log$F CaII H  & $\log$F CaII K  & $\log$F H$\alpha$  & $\log$F H$\beta$  & $\log$F H$\gamma$  & $\log$F H$\delta$  & $\log$F H$\epsilon$  & $\log$F$_{\rm X}$  \\ 
 (1)    &   (2)         &  (3)       & (4)      &   (5)            & (6)             &  (7)           & (8)           &  (9) \\
\hline
\endfirsthead
\caption{Continued.} \\
\hline
 Star & $\log$F CaII H  & $\log$F CaII K  & $\log$F H$\alpha$  & $\log$F H$\beta$  & $\log$F H$\gamma$  & $\log$F H$\delta$  & $\log$F H$\epsilon$  & $\log$F$_{\rm X}$  \\
 (1)    &   (2)         &  (3)       & (4)      &   (5)            & (6)             &  (7)           & (8)           &  (9) \\ 
\hline
\endhead
\hline
\endfoot
\hline
\endlastfoot
GJ 2	&	4.944	$\pm$	0.002	&	5.020	$\pm$	0.001	&	4.685	$\pm$	0.011	&	4.531	$\pm$	0.016	&	4.091	$\pm$	0.014	&	3.849	$\pm$	0.028	&	3.850	$\pm$	0.020	&				\\
GJ 3014	&	5.172	$\pm$	0.032	&	5.142	$\pm$	0.034	&	5.036	$\pm$	0.014	&	4.787	$\pm$	0.028	&	4.386	$\pm$	0.083	&	4.308	$\pm$	0.155	&	4.330	$\pm$	0.220	&	5.82	$\pm$	0.17	\\
GJ 16	&	4.817	$\pm$	0.003	&	4.881	$\pm$	0.002	&	3.736	$\pm$	0.107	&	4.289	$\pm$	0.031	&	3.424	$\pm$	0.061	&	3.182	$\pm$	0.135	&	3.720	$\pm$	0.033	&				\\
GJ 15A	&				&	2.457	$\pm$	0.188	&	4.059	$\pm$	0.020	&	3.770	$\pm$	0.031	&	3.251	$\pm$	0.035	&				&				&	5.35	$\pm$	0.12	\\
GJ 21	&	5.093	$\pm$	0.002	&	5.165	$\pm$	0.002	&	4.952	$\pm$	0.006	&	4.646	$\pm$	0.010	&	4.214	$\pm$	0.013	&	4.137	$\pm$	0.019	&	4.110	$\pm$	0.019	&				\\
GJ 26	&	4.577	$\pm$	0.006	&	4.627	$\pm$	0.005	&				&				&	3.344	$\pm$	0.175	&	3.555	$\pm$	0.103	&	3.601	$\pm$	0.054	&	5.27	$\pm$	0.23	\\
GJ 47	&	4.569	$\pm$	0.006	&	4.621	$\pm$	0.006	&				&	4.057	$\pm$	0.051	&	3.816	$\pm$	0.042	&	3.793	$\pm$	0.041	&	3.701	$\pm$	0.047	&	5.35	$\pm$	0.31	\\
GJ 49	&	5.194	$\pm$	0.001	&	5.258	$\pm$	0.001	&	5.170	$\pm$	0.005	&	4.877	$\pm$	0.010	&	4.437	$\pm$	0.008	&	4.248	$\pm$	0.015	&	4.216	$\pm$	0.009	&				\\
GJ 1030	&	4.897	$\pm$	0.011	&	4.918	$\pm$	0.010	&	3.741	$\pm$	0.168	&	4.333	$\pm$	0.041	&	3.540	$\pm$	0.130	&	2.722	$\pm$	1.334	&	3.869	$\pm$	0.114	&				\\
NLTT 4188	&	5.085	$\pm$	0.021	&	5.075	$\pm$	0.021	&	4.908	$\pm$	0.013	&	4.490	$\pm$	0.033	&	4.350	$\pm$	0.058	&	4.276	$\pm$	0.085	&	4.347	$\pm$	0.115	&				\\
GJ 70	&	4.627	$\pm$	0.006	&	4.661	$\pm$	0.006	&	3.670	$\pm$	0.167	&	4.251	$\pm$	0.035	&	3.955	$\pm$	0.038	&	3.979	$\pm$	0.036	&	3.790	$\pm$	0.043	&	5.47	$\pm$	0.14	\\
GJ 3117A	&	4.790	$\pm$	0.009	&	4.859	$\pm$	0.008	&				&	4.244	$\pm$	0.039	&	3.326	$\pm$	0.162	&	3.525	$\pm$	0.141	&	3.779	$\pm$	0.092	&				\\
GJ 3126	&	4.461	$\pm$	0.023	&	4.562	$\pm$	0.018	&				&				&				&				&	3.693	$\pm$	0.133	&				\\
GJ 3186	&	4.890	$\pm$	0.024	&	4.963	$\pm$	0.020	&	4.444	$\pm$	0.043	&	4.393	$\pm$	0.041	&	3.560	$\pm$	0.258	&				&	3.909	$\pm$	0.231	&				\\
GJ 119A	&	4.392	$\pm$	0.008	&	4.434	$\pm$	0.008	&				&	3.801	$\pm$	0.145	&				&				&	3.532	$\pm$	0.060	&				\\
GJ 119B	&	4.473	$\pm$	0.030	&	4.635	$\pm$	0.021	&				&				&				&				&	2.337	$\pm$	4.164	&				\\
TYC 1795-941-1	&	5.001	$\pm$	0.032	&	5.057	$\pm$	0.028	&	3.377	$\pm$	0.887	&				&				&	4.259	$\pm$	0.122	&	4.094	$\pm$	0.261	&				\\
NLTT 10614	&	4.688	$\pm$	0.061	&	4.848	$\pm$	0.042	&	4.557	$\pm$	0.045	&	3.991	$\pm$	0.134	&	3.789	$\pm$	0.216	&				&	3.450	$\pm$	1.056	&				\\
TYC 3720-426-1	&	5.775	$\pm$	0.004	&	5.831	$\pm$	0.004	&	6.163	$\pm$	0.001	&	5.737	$\pm$	0.004	&	5.472	$\pm$	0.005	&	5.430	$\pm$	0.006	&	5.267	$\pm$	0.013	&	6.79	$\pm$	0.18	\\
GJ 150.1B	&	4.982	$\pm$	0.003	&	5.057	$\pm$	0.002	&	4.823	$\pm$	0.009	&	4.592	$\pm$	0.016	&	4.048	$\pm$	0.020	&	3.922	$\pm$	0.032	&	4.018	$\pm$	0.027	&				\\
GJ 156.1A	&	4.788	$\pm$	0.005	&	4.845	$\pm$	0.005	&	4.257	$\pm$	0.070	&	4.184	$\pm$	0.083	&				&	2.524	$\pm$	1.201	&	3.662	$\pm$	0.069	&				\\
GJ 162	&	4.923	$\pm$	0.002	&	5.005	$\pm$	0.002	&	4.694	$\pm$	0.008	&	4.412	$\pm$	0.015	&	3.807	$\pm$	0.036	&	3.826	$\pm$	0.036	&	3.736	$\pm$	0.038	&				\\
GJ 1074	&	4.796	$\pm$	0.007	&	4.875	$\pm$	0.006	&	3.875	$\pm$	0.058	&	4.076	$\pm$	0.034	&	3.496	$\pm$	0.087	&	3.125	$\pm$	0.257	&	3.430	$\pm$	0.158	&				\\
GJ 184	&	3.976	$\pm$	0.025	&	4.072	$\pm$	0.020	&				&	3.595	$\pm$	0.169	&				&	3.297	$\pm$	0.118	&	3.437	$\pm$	0.087	&				\\
GJ 3352	&	4.932	$\pm$	0.014	&	5.017	$\pm$	0.012	&	4.730	$\pm$	0.017	&	4.226	$\pm$	0.040	&	3.719	$\pm$	0.124	&	4.059	$\pm$	0.075	&	3.803	$\pm$	0.194	&				\\
TYC 3379-1077-1	&	5.029	$\pm$	0.039	&	5.139	$\pm$	0.030	&	4.848	$\pm$	0.019	&				&	3.650	$\pm$	0.497	&				&	4.130	$\pm$	0.309	&				\\
TYC 743-1836-1	&	5.188	$\pm$	0.017	&	5.262	$\pm$	0.014	&	5.194	$\pm$	0.010	&	4.576	$\pm$	0.040	&	4.376	$\pm$	0.056	&	4.567	$\pm$	0.047	&	4.323	$\pm$	0.123	&				\\
GJ 272	&	4.888	$\pm$	0.007	&	4.904	$\pm$	0.007	&				&	4.095	$\pm$	0.048	&				&	2.734	$\pm$	0.788	&	3.412	$\pm$	0.206	&	5.34	$\pm$	0.27	\\
StKM 1-650	&	4.967	$\pm$	0.023	&	4.998	$\pm$	0.021	&	4.975	$\pm$	0.010	&	4.279	$\pm$	0.054	&	4.087	$\pm$	0.072	&	2.940	$\pm$	1.588	&	4.167	$\pm$	0.145	&				\\
NLTT 21156	&	5.308	$\pm$	0.003	&	5.374	$\pm$	0.003	&	5.544	$\pm$	0.003	&	5.226	$\pm$	0.004	&	4.935	$\pm$	0.004	&	4.785	$\pm$	0.008	&	4.643	$\pm$	0.014	&	6.02	$\pm$	0.28	\\
GJ 399	&	4.418	$\pm$	0.014	&	4.533	$\pm$	0.011	&				&	4.074	$\pm$	0.064	&	3.195	$\pm$	0.181	&				&	3.241	$\pm$	0.216	&				\\
GJ 408	&	4.550	$\pm$	0.005	&	4.610	$\pm$	0.004	&				&				&	3.289	$\pm$	0.207	&	3.346	$\pm$	0.161	&	3.375	$\pm$	0.072	&	5.06	$\pm$	0.26	\\
GJ 412A	&				&				&	4.667	$\pm$	0.006	&	4.039	$\pm$	0.020	&	3.704	$\pm$	0.020	&	3.266	$\pm$	0.051	&	2.432	$\pm$	0.239	&	5.46	$\pm$	0.13	\\
GJ 414B	&	4.771	$\pm$	0.003	&	4.840	$\pm$	0.003	&	4.295	$\pm$	0.040	&	4.358	$\pm$	0.036	&	3.464	$\pm$	0.081	&	2.837	$\pm$	0.411	&	3.623	$\pm$	0.046	&				\\
GJ 3649	&	4.979	$\pm$	0.006	&	5.048	$\pm$	0.005	&	4.838	$\pm$	0.011	&	4.605	$\pm$	0.018	&	4.114	$\pm$	0.031	&	4.022	$\pm$	0.042	&	4.032	$\pm$	0.053	&				\\
GJ 450	&	4.979	$\pm$	0.002	&	5.060	$\pm$	0.002	&	5.144	$\pm$	0.003	&	4.875	$\pm$	0.006	&	4.554	$\pm$	0.005	&	4.461	$\pm$	0.007	&	4.317	$\pm$	0.008	&	5.56	$\pm$	0.17	\\
GJ 9404	&	4.898	$\pm$	0.005	&	4.973	$\pm$	0.004	&	4.707	$\pm$	0.017	&	3.941	$\pm$	0.108	&	3.194	$\pm$	0.271	&	3.730	$\pm$	0.083	&	3.864	$\pm$	0.057	&	5.56	$\pm$	0.12	\\
GJ 476	&	4.358	$\pm$	0.018	&	4.447	$\pm$	0.015	&				&				&				&				&	2.593	$\pm$	1.068	&	5.11	$\pm$	0.15	\\
GJ 9440	&	4.709	$\pm$	0.005	&	4.791	$\pm$	0.004	&				&	4.212	$\pm$	0.054	&				&				&	3.310	$\pm$	0.115	&	5.02	$\pm$	0.09	\\
GJ 521A	&	4.465	$\pm$	0.004	&	4.488	$\pm$	0.004	&				&				&				&				&	3.319	$\pm$	0.054	&				\\
GJ 3822	&	5.136	$\pm$	0.003	&	5.196	$\pm$	0.002	&	5.043	$\pm$	0.006	&	4.659	$\pm$	0.012	&	4.166	$\pm$	0.020	&	4.181	$\pm$	0.025	&	4.194	$\pm$	0.023	&	5.50	$\pm$	0.26	\\
GJ 548A	&	5.008	$\pm$	0.003	&	5.066	$\pm$	0.003	&	4.950	$\pm$	0.010	&	4.247	$\pm$	0.056	&	3.826	$\pm$	0.063	&	4.043	$\pm$	0.039	&	4.117	$\pm$	0.027	&	5.69	$\pm$	0.17	\\
GJ 552	&	4.624	$\pm$	0.004	&	4.707	$\pm$	0.003	&				&	4.164	$\pm$	0.047	&	3.465	$\pm$	0.072	&	3.394	$\pm$	0.107	&	3.639	$\pm$	0.037	&				\\
GJ 606	&	4.956	$\pm$	0.005	&	5.017	$\pm$	0.004	&	4.890	$\pm$	0.009	&	4.642	$\pm$	0.013	&	4.236	$\pm$	0.015	&	4.082	$\pm$	0.031	&	4.023	$\pm$	0.040	&	5.64	$\pm$	0.28	\\
GJ 3942	&	5.178	$\pm$	0.002	&	5.248	$\pm$	0.002	&	5.092	$\pm$	0.004	&	4.621	$\pm$	0.010	&	4.200	$\pm$	0.016	&	4.239	$\pm$	0.017	&	4.264	$\pm$	0.016	&	5.33	$\pm$	0.19	\\
GJ 625	&	4.259	$\pm$	0.008	&	4.354	$\pm$	0.007	&				&				&	3.128	$\pm$	0.217	&	3.590	$\pm$	0.063	&	3.010	$\pm$	0.148	&	5.11	$\pm$	0.17	\\
GJ 3997	&	-9.000	$\pm$	-9.000	&	2.658	$\pm$	0.381	&				&	2.649	$\pm$	0.891	&	3.016	$\pm$	0.193	&	2.995	$\pm$	0.216	&	2.806	$\pm$	0.271	&				\\
GJ 3998	&	4.725	$\pm$	0.004	&	4.796	$\pm$	0.004	&				&	4.098	$\pm$	0.051	&				&				&	3.536	$\pm$	0.068	&				\\
GJ 2128	&	4.025	$\pm$	0.045	&	4.083	$\pm$	0.039	&				&				&				&				&	&				\\
GJ 671	&	4.278	$\pm$	0.020	&	4.259	$\pm$	0.021	&				&				&	3.583	$\pm$	0.128	&				&				&				\\
GJ 685	&	5.083	$\pm$	0.002	&	5.149	$\pm$	0.002	&	4.889	$\pm$	0.007	&	4.529	$\pm$	0.013	&	3.971	$\pm$	0.026	&	4.082	$\pm$	0.025	&	4.139	$\pm$	0.021	&	5.28	$\pm$	0.11	\\
GJ 686	&	3.601	$\pm$	0.096	&	3.661	$\pm$	0.084	&	4.072	$\pm$	0.038	&	4.039	$\pm$	0.036	&	3.125	$\pm$	0.211	&				&	3.542	$\pm$	0.110	&				\\
GJ 694.2	&	4.819	$\pm$	0.005	&	4.891	$\pm$	0.004	&	4.718	$\pm$	0.008	&	4.250	$\pm$	0.019	&	3.529	$\pm$	0.060	&	3.758	$\pm$	0.045	&	3.928	$\pm$	0.037	&	5.22	$\pm$	0.23	\\
GJ 4057	&	4.630	$\pm$	0.008	&	4.667	$\pm$	0.007	&	4.251	$\pm$	0.025	&	3.606	$\pm$	0.110	&	2.484	$\pm$	0.714	&	3.536	$\pm$	0.085	&	3.728	$\pm$	0.062	&				\\
GJ 720A	&	3.942	$\pm$	0.021	&	3.935	$\pm$	0.021	&				&				&				&	2.649	$\pm$	0.451	&	2.972	$\pm$	0.192	&	5.12	$\pm$	0.18	\\
GJ 731	&	4.764	$\pm$	0.009	&	4.809	$\pm$	0.008	&	4.792	$\pm$	0.007	&	4.294	$\pm$	0.019	&	3.948	$\pm$	0.034	&	4.051	$\pm$	0.033	&	4.024	$\pm$	0.050	&				\\
GJ 740	&	4.755	$\pm$	0.004	&	4.825	$\pm$	0.004	&	4.364	$\pm$	0.020	&	3.944	$\pm$	0.046	&				&	3.347	$\pm$	0.123	&	3.655	$\pm$	0.053	&	5.24	$\pm$	0.20	\\
GJ 4092	&	4.321	$\pm$	0.029	&	4.278	$\pm$	0.032	&				&				&				&	3.355	$\pm$	0.262	&	3.380	$\pm$	0.253	&				\\
GJ 9689	&	4.745	$\pm$	0.010	&	4.813	$\pm$	0.009	&	4.311	$\pm$	0.032	&	3.955	$\pm$	0.069	&				&	2.867	$\pm$	0.606	&	3.738	$\pm$	0.103	&				\\
GJ 793	&	4.802	$\pm$	0.003	&	4.880	$\pm$	0.003	&	4.200	$\pm$	0.068	&	4.540	$\pm$	0.026	&	4.415	$\pm$	0.016	&	4.248	$\pm$	0.021	&	4.077	$\pm$	0.017	&	5.92	$\pm$	0.18	\\
BPM 96441	&	5.015	$\pm$	0.014	&	5.116	$\pm$	0.011	&	4.878	$\pm$	0.018	&	3.258	$\pm$	0.846	&	3.675	$\pm$	0.184	&	4.192	$\pm$	0.071	&	4.054	$\pm$	0.127	&				\\
TYC 2710-691-1	&	5.203	$\pm$	0.021	&	5.245	$\pm$	0.019	&	5.061	$\pm$	0.019	&	4.471	$\pm$	0.077	&	4.524	$\pm$	0.058	&	4.515	$\pm$	0.084	&	4.395	$\pm$	0.138	&	6.34	$\pm$	0.29	\\
TYC 2703-706-1	&	5.854	$\pm$	0.001	&	5.893	$\pm$	0.001	&	6.251	$\pm$	0.001	&	5.919	$\pm$	0.001	&	5.647	$\pm$	0.001	&	5.541	$\pm$	0.002	&	5.379	$\pm$	0.003	&	6.60	$\pm$	0.27	\\
GJ 4196	&	3.144	$\pm$	1.207	&				&	3.840	$\pm$	0.222	&	4.076	$\pm$	0.091	&				&				&	3.349	$\pm$	0.752	&				\\
NLTT 52021	&	4.970	$\pm$	0.035	&	5.090	$\pm$	0.026	&	4.711	$\pm$	0.025	&	4.537	$\pm$	0.041	&	3.459	$\pm$	0.480	&	3.831	$\pm$	0.318	&	3.393	$\pm$	1.311	&				\\
NLTT 53166	&	4.449	$\pm$	0.026	&	4.601	$\pm$	0.018	&	4.175	$\pm$	0.043	&				&	3.431	$\pm$	0.177	&	3.876	$\pm$	0.093	&	3.633	$\pm$	0.171	&				\\
2MASS J22353504+3712131	&	5.096	$\pm$	0.022	&	5.177	$\pm$	0.018	&	5.002	$\pm$	0.015	&	3.874	$\pm$	0.219	&	4.120	$\pm$	0.099	&	4.484	$\pm$	0.066	&	4.206	$\pm$	0.170	&				\\
GJ 9793 	&	5.184	$\pm$	0.003	&	5.230	$\pm$	0.003	&	5.545	$\pm$	0.003	&	4.980	$\pm$	0.021	&	4.759	$\pm$	0.016	&	4.762	$\pm$	0.013	&	4.628	$\pm$	0.011	&	6.29	$\pm$	0.35	\\
GJ 4306	&	4.983	$\pm$	0.002	&	5.039	$\pm$	0.002	&	4.654	$\pm$	0.011	&	4.433	$\pm$	0.021	&	3.633	$\pm$	0.042	&	3.737	$\pm$	0.042	&	3.953	$\pm$	0.021	&				\\
GJ 895	&	4.853	$\pm$	0.004	&	4.920	$\pm$	0.003	&	4.378	$\pm$	0.031	&	4.376	$\pm$	0.032	&	3.465	$\pm$	0.101	&	3.333	$\pm$	0.152	&	3.695	$\pm$	0.057	&	4.98	$\pm$	0.27	\\
V* BRPsc	&	4.015	$\pm$	0.007	&	3.975	$\pm$	0.008	&				&				&				&	1.921	$\pm$	1.713	&	3.022	$\pm$	0.069	&	5.24	$\pm$	0.17	\\
\end{longtable}
\end{landscape}
\end{longtab}
}

\subsection{Relationships between rotation, activity, and stellar parameters}
\subsubsection{Activity versus effective temperature}
  
\begin{figure*}[!htb]
\centering
\includegraphics[angle=270,scale=0.54]{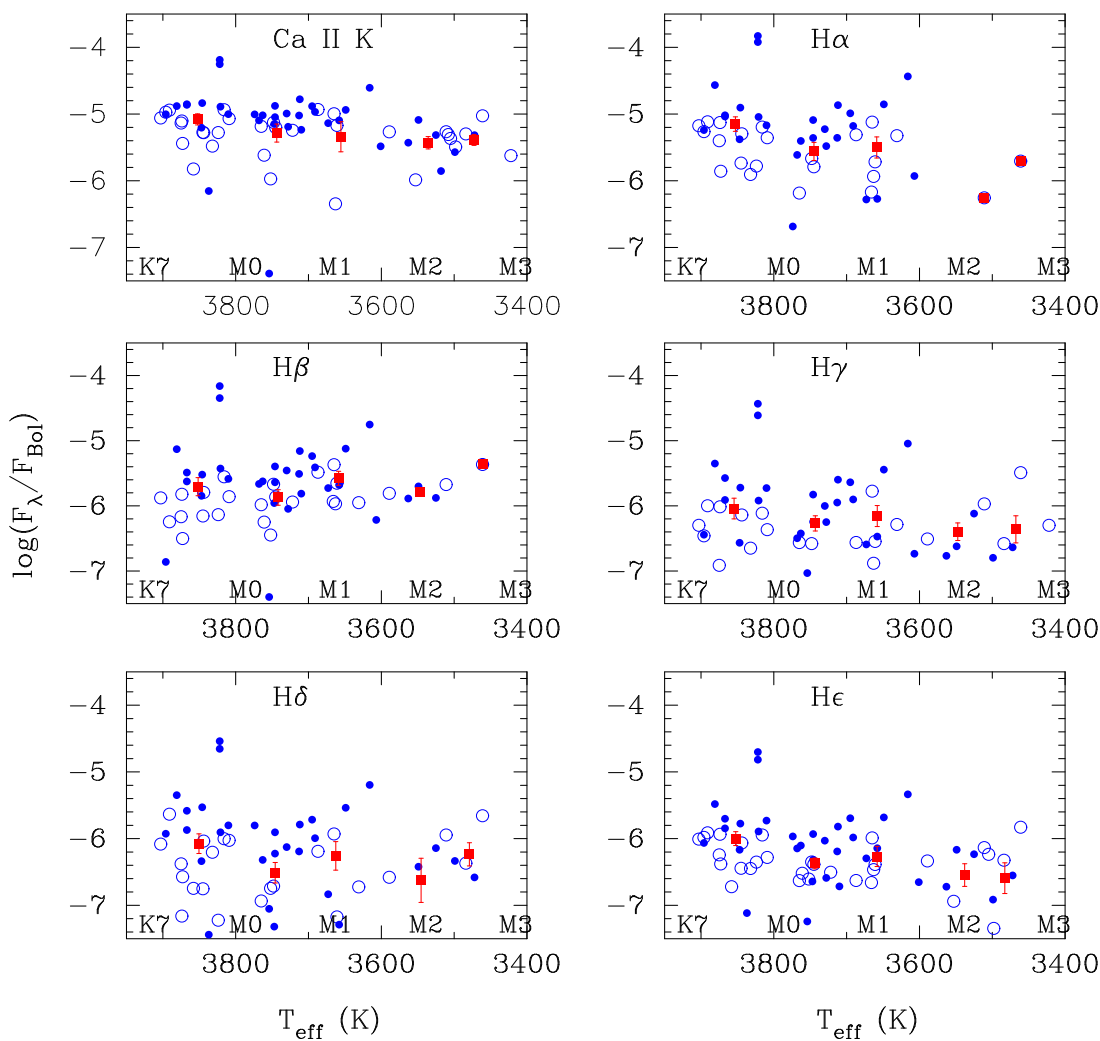}
\caption{
$\log($F$_{\lambda}$/F$_{\rm Bol})$ values
vs. the stellar effective temperature (K).
Possible young stars  according to our kinematic analysis are shown with filled symbols.  
Median binned values are overplotted as red squares.
The relationship between T$_{\rm eff}$ and spectral type
is taken from \cite{2015A&A...577A.132M}.
}
\label{halpha_teff}
\end{figure*}

  Figure~\ref{halpha_teff} shows the excess 
  $\log($F$_{\lambda}$/F$_{\rm Bol})$ values
  as a function of the stellar effective temperature,
  where bolometric fluxes 
  were computed from the stellar luminosities and radii given in 
  Table~\ref{parameters_table_full}.
  Besides a large
  scatter, the strength of the emission excess 
  is roughly constant for the stars in the temperature
  range 
  studied here (3400-3980 K, spectral types K7.5-M3).
  This result holds for all the considered activity indicators.
  This is in line with previous works showing that the strength of the 
  activity is constant for early-mid M types
  \citep{1996AJ....112.2799H,2004AJ....128..426W,2008PASP..120.1161W,
  2012AJ....143...93R,2013A&A...558A.141S}.
  The comparison with the literature samples also reveals
  the low-activity levels of our sample.
  While most of our stars show values of $\log($F$_{H\alpha}$/F$_{\rm Bol})$
  lower than -5.0, the values in the literature are tipically in the range between
 -3.0 and -4.5.  

  We also note that the median $\log$F$_{\lambda}$ values
  are higher in the calcium lines.
  It also tends to decrease through the Balmer lines towards lower wavelengths,
  from  4.72 [erg cm$^{\rm -2}$ s$^{\rm -1}$]  for H${\alpha}$ to 3.78 
  [erg cm$^{\rm -2}$ s$^{\rm -1}$] for H${\epsilon}$.
  This fact is likely related to the different height/regions
  in which the lines are formed. 
 
\subsubsection{Rotational velocity versus effective temperature}
\label{rotteff}

\begin{figure}
\centering
\includegraphics[angle=270,scale=0.45]{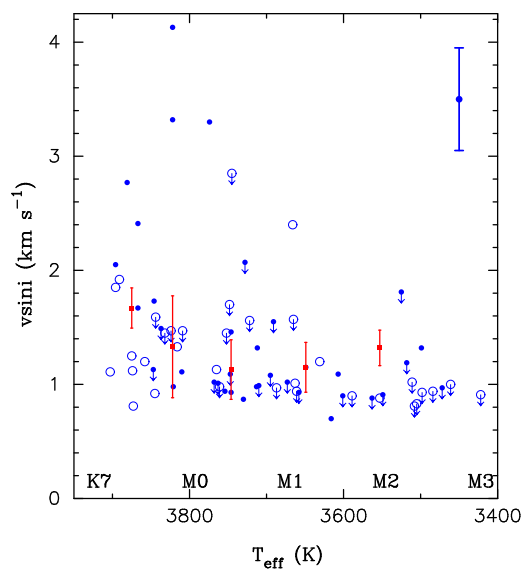}
\caption{
v$\sin i$ vs. the effective temperature.
Upper limits on  v$\sin i$ are shown with arrows. 
Possible young stars are shown with filled symbols.
Typical uncertainties are also shown.
Median binned values (without considering upper limits) are overplotted as red squares.
}
\label{vsini_vs_teff}
\end{figure}

 We next explore the correlation between rotation and effective temperature.
 Figure~\ref{vsini_vs_teff} shows the derived v$\sin i$ values as a function
 of T$_{\rm eff}$. Several conclusions can be drawn from this figure.
 First, we note the very low rotation levels of
 our sample. Only four stars show v$\sin i$ values larger than
 2.5 km s$^{\rm -1}$, namely
 GJ 9793, TYC 1795-941-1, TYC 3720-426-1,
 and TYC 2703-706-1.
 Although our sample might be biased towards slow rotating stars,
 our analysis is in agreement with previous results suggesting that rotation levels
 larger than about 2.5 km s$^{\rm -1}$ are rare in old field M stars
 \citep[e.g.][]{1992ApJ...390..550M,2010AJ....139..504B}.

 Figure~\ref{vsini_vs_teff} also shows that our capability to measure v$\sin i$ severely
 diminishes as we move towards cooler stars. In particular, for stars
 cooler than 3700 K,  $\sim$ 72\% of the measurements correspond to upper limits.
 Furthermore, our estimated uncertainties slightly increase towards cooler stars.
 We also note that there seems to be a tendency to lower rotation
 levels as we move towards cooler stars. This is especially evident
 if we look at the binned v$\sin i$ values (red squares in
 Figure~\ref{vsini_vs_teff}).
 These two effects may be related. We should first remind that the errors
 in v$\sin i$ tend to increase towards lower  v$\sin i$ values
 and it could be that errors on  v$\sin i$ increase towards cooler stars
 just because the v$\sin i$ diminishes towards lower temperatures.
 This fact reflects a relation between low rotation rates and late-type stars
 as well as the difficulty in deriving accurate v$\sin i$ values
 for these stars. 

 In order to test whether a temperature-rotation correlation is present
 in our data several statistical tests were performed:
 {\it i)} the Spearman's correlation test excluding upper limits;
 and {\it ii)} the generalised Kendall's $\tau$ and generalised Spearman's $\rho$ correlation tests.
 The latest were performed using the ASURV code \citep{2014ascl.soft06001F}, which implements the
 methods presented in \cite{1986ApJ...306..490I}.
 Table~\ref{asurv_statistics} shows the results.
 While the Spearman's test (excluding upper limits) suggests that there is no
 correlation, the results from the generalised Kendall's and  Spearman's 
 tests are compatible with a moderate but statistically significant trend of 
 lower rotation levels towards cooler stars.
  This result, however, should be regarded with caution as it could be the
 case that the kinematically old stars in our sample are cooler than
 the possible young ones. While a K-S test shows no significant differences
 in the T$_{\rm eff}$  distribution of kinematically young/old stars
 ($D$ $\sim$ 0.15, $p$ $\sim$ 0.77), a tendency of slightly lower fraction
 of kinematically young stars towards cooler temperatures might be present
 in our data (45.8\%, 66.7\%, 57.1\%, 44.4\%, and 33.3\%, for stars in
 the  T$_{\rm eff}$ ranges: 3800-3900 K, 3700-3800 K, 3600-3700 K, 3500-3400 K,
 3500-3600 K, and 3400-3500 K, respectively).

 If confirmed, the trend of lower rotation levels towards cooler stars
 might appear to be in contradiction
 with \cite{2010AJ....139..504B} who found that the fraction of
 stars with v$\sin i$ larger 
 than 2.5 km s$^{\rm -1}$ increases towards lower masses.
 However, we note that these authors consider stars with spectral
 types from M0 to $\sim$ M6, while our sample is limited 
 to M3. In particular, the rise in the fraction of stars with
 v$\sin i$  $>$ 2.5 km s$^{\rm -1}$ noted in \cite{2010AJ....139..504B} seems to start 
 at spectral types around  M3/M3.5 i.e., corresponding to the transition between
 partially and fully convective stars where the values and the spread on the v$\sin i$ are known
 to be large \citep[see for e.g.][]{2012AJ....143...93R,2013MNRAS.431.2063S}.
 For hotter stars, our results do not
 seem to differ from \citet[][Fig.~2]{2010AJ....139..504B}.  
 Further, although the sample of \cite{2010AJ....139..504B} was
 selecting for radial velocity monitoring, no selection
 of low-activity targets was made. Indeed, the authors cautioned
 that their sample may be biased towards nearby and implicitly young
 (and somewhat more rapidly rotating) targets.
 These results are also in line with the findings by \cite{1998A&A...331..581D}
 who found no measurable rotation for stars in the range M0-M3, whist they 
 identified an increasing fraction of rotating stars among their sample
 of dynamical young stars for spectral types above M3.

\subsubsection{The rotational velocity - activity relationship}
\begin{figure*}[!htb]
\centering
\begin{minipage}{0.49\linewidth}
\includegraphics[angle=270,scale=0.45]{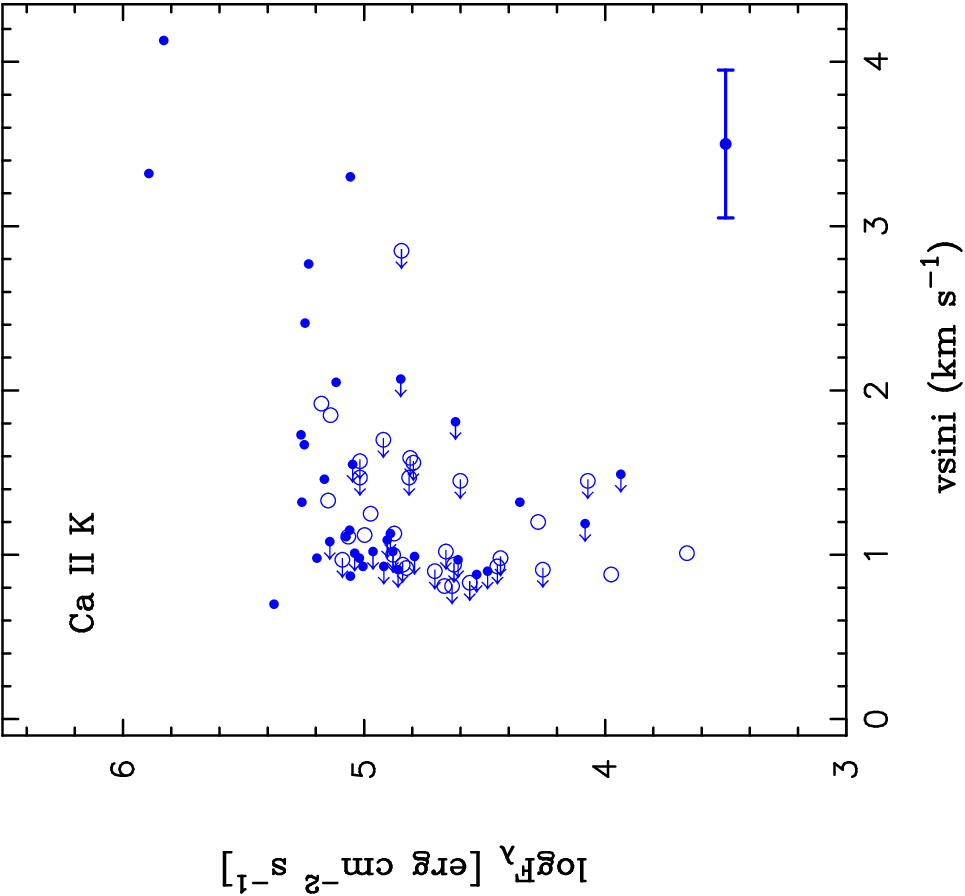}
\end{minipage}
\begin{minipage}{0.49\linewidth}
\includegraphics[angle=270,scale=0.45]{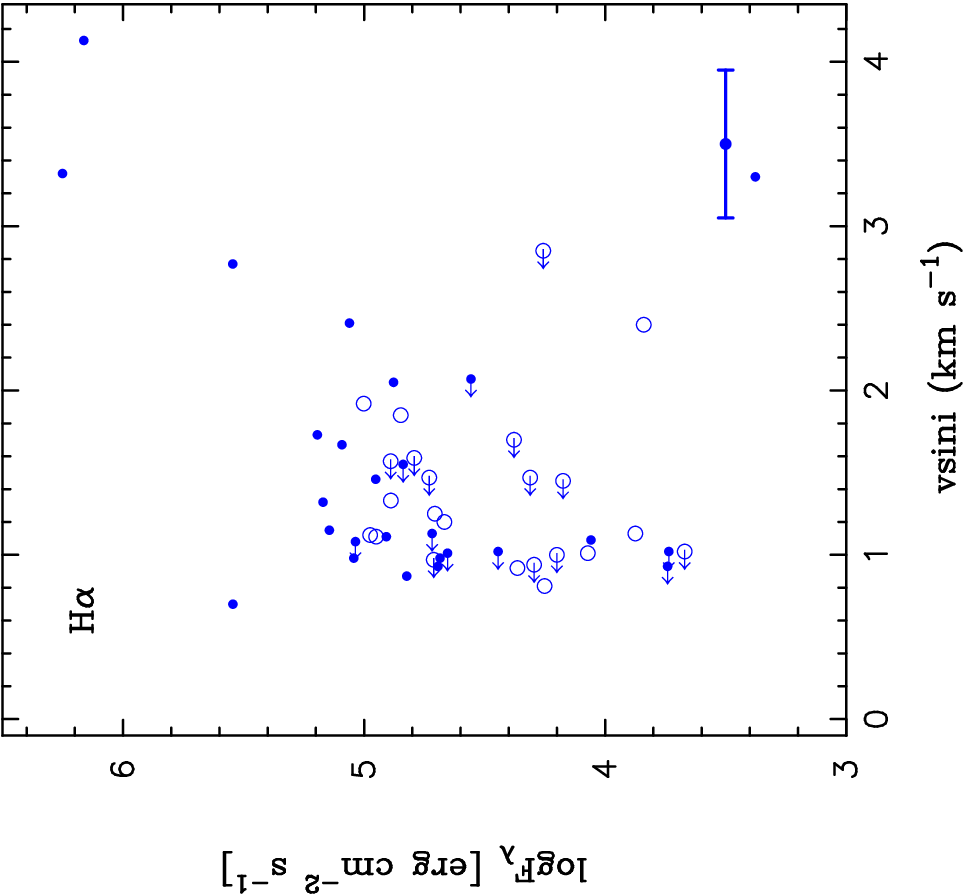}
\end{minipage}
\caption{
Ca~{\sc ii} K (left) and H${\alpha}$ (right) line
emission excess flux 
vs. v$\sin i$.
Possible young stars according to our kinematic analysis are shown with filled symbols.
Upper limits on  v$\sin i$ are shown with arrows,
while the typical error bar is shown on the lower right.} 
\label{vsini_vs_caiik}
\end{figure*}

 Figure~\ref{vsini_vs_caiik} shows 
 the flux excess of the Ca~{\sc ii K} (left) and H${\alpha}$ (right) emission 
 as a function of our measured v$\sin i$ values. 
 The figure reveals for both lines a mild tendency of higher v$\sin i$ values with increasing
 activity strength. 
 We should caution that several factors may be 
 affecting our study. 

 To start with, we are considering
 v$\sin i$ values and not rotational periods
 so the $\sin i$ term introduces additional scatter
 and long rotation periods are not covered\footnote{For a typical radius of 0.5 R$_{\odot}$ and assuming
 a low v$\sin i$ of 0.8 km s$^{\rm -1}$, 
 rotation periods  longer than $\sim$
 32 days are excluded from the analysis.} \citep[e.g.][]{2015ApJ...812....3W}.
 Further, our sample is limited
 to a narrow range in spectral type. Furthermore, our sample 
 is also biased towards
 low-rotation and low-activity stars.
 As in the previous subsection, several statistical tests were performed.
 Table~\ref{asurv_statistics} shows the results for the Ca~{\sc ii K} and H${\alpha}$ line.
 We conclude that the statistical tests
 suggest that a moderate but significant correlation between activity
 and rotation might be present 
 present in our data. We also note that both the strength and the statistical
 significance of the correlation are higher in the Ca~{\sc ii K} than in the H${\alpha}$ line.

 We are aware that the visual inspection of Figure~\ref{vsini_vs_caiik}
  does not clearly support the results from the statistical tests as the majority
  of the stars show projected rotational velocities on the 1-2 kms$^{\rm -1}$ range,
  most show upper limits on v$\sin i$, and the range of rotational velocities considered here is
  rather small (up to 4 kms$^{\rm -1}$).
  In order to check the statistical significance of the possible activity-rotation correlation, 
  all the statistical tests were repeated excluding the stars with the highest
  rotational velocities (v$\sin i$ $>$ 2.5 kms$^{\rm -1}$). In this way we can check
  whether the correlation is (or not) dominated by these few stars.
  The results are given in Table~\ref{asurv_statistics2}. As expected, both the strength and the statistical
  significance of the correlation clearly diminishes.
  However, the statistical significance of the correlations 
  (for the analysis including the upper limits) are still well beyond 
  98\%  (the typical admitted threshold for considering statistical significance). 

  We conclude that a hint of a rotation-activity correlation might be present in
  our data, although the analysis of larger samples of stars covering a wider
  range of rotation and
  with measured photometric periods might be needed to clearly confirm it.

\begin{table}
\centering
\caption{
Results from Spearman's correlation test
and the generalised Kendall's $\tau$ and  Spearman's $\rho$ correlation tests.
}
\label{asurv_statistics}
\begin{tabular}{lll}
\hline
\multicolumn{3}{c}{v$\sin i$ vs. T$_{\rm eff}$}\\
\hline
Spearman's test$^{\dag}$ & $\rho$    $\sim$ 0.37 & $p$ $\sim$ 0.032       \\
Generalised Kendall      & $Z$       $\sim$ 4.53 & $p$ $<$ 10$^{\rm -4}$ \\
Generalised Spearman     & $\rho$    $\sim$ 0.57 & $p$ $<$ 10$^{\rm -4}$ \\
\hline
\multicolumn{3}{c}{$\log$F Ca~{\sc ii}  K vs. v$\sin i$}\\
\hline
Spearman's test$^{\dag}$ & $\rho$    $\sim$ 0.55 & $p$ 0.001 \\
Generalised Kendall      & $Z$       $\sim$ 4.60 & $p$ $<$ 10$^{\rm -4}$ \\
Generalised Spearman     & $\rho$    $\sim$ 0.53 & $p$ $<$ 10$^{\rm -4}$ \\
\hline
\multicolumn{3}{c}{$\log$F H$\alpha$ vs. v$\sin i$ }\\
\hline
Spearman's test$^{\dag}$ & $\rho$ $\sim$ 0.33 & $p$ $\sim$  0.079 \\
Generalised Kendall      & $Z$    $\sim$ 3.02 & $p$ $\sim$  0.003 \\
Generalised Spearman     & $\rho$ $\sim$ 0.43 & $p$ $\sim$  0.003 \\ 
\hline
\multicolumn{3}{c}{$\log$F X-ray vs. v$\sin i$ }\\
\hline
Spearman's test$^{\dag}$ & $\rho$ $\sim$ 0.48 & $p$ $\sim$  0.058 \\
Generalised Kendall      & $Z$    $\sim$ 2.64 & $p$ $\sim$  0.008 \\
Generalised Spearman     & $\rho$ $\sim$ 0.43 & $p$ $\sim$  0.022 \\
\hline
\end{tabular}
\tablefoot{ $^{\dag}$ Upper limits excluded.} 
\end{table}

\begin{table}
\centering
\caption{
Results from Spearman's correlation test 
and the generalised Kendall's $\tau$ and  Spearman's $\rho$ correlation tests,
excluding those stars with v$\sin i$ $>$ 2.5 kms$^{\rm -1}$.
}
\label{asurv_statistics2}
\begin{tabular}{lll}
\hline
\multicolumn{3}{c}{$\log$F Ca~{\sc ii}  K vs. v$\sin i$}\\
\hline
Spearman's test$^{\dag}$ & $\rho$    $\sim$ 0.45 & $p$ $\sim$ 0.0155 \\
Generalised Kendall      & $Z$       $\sim$ 3.66 & $p$ $\sim$ 0.0003 \\
Generalised Spearman     & $\rho$    $\sim$ 0.45 & $p$ $\sim$ 0.0004 \\
\hline
\multicolumn{3}{c}{$\log$F H$\alpha$ vs. v$\sin i$ }\\
\hline
Spearman's test$^{\dag}$ & $\rho$ $\sim$ 0.22 & $p$ $\sim$  0.2720 \\
Generalised Kendall      & $Z$    $\sim$ 2.51 & $p$ $\sim$  0.0119 \\
Generalised Spearman     & $\rho$ $\sim$ 0.41 & $p$ $\sim$  0.0072 \\
\hline
\end{tabular}
\end{table}

\subsubsection{Age effects}

  Starting approx. from the zero age main-sequence (ZAMS),
  stellar activity and rotation are expected to decrease with time
  as a star loses angular momentum with stellar winds  via
  magnetic braking \citep{1967ApJ...148..217W,1993MNRAS.261..766J}. 

\begin{figure}[!htb]
\centering
\begin{minipage}{0.49\linewidth}
\includegraphics[angle=270,scale=0.45]{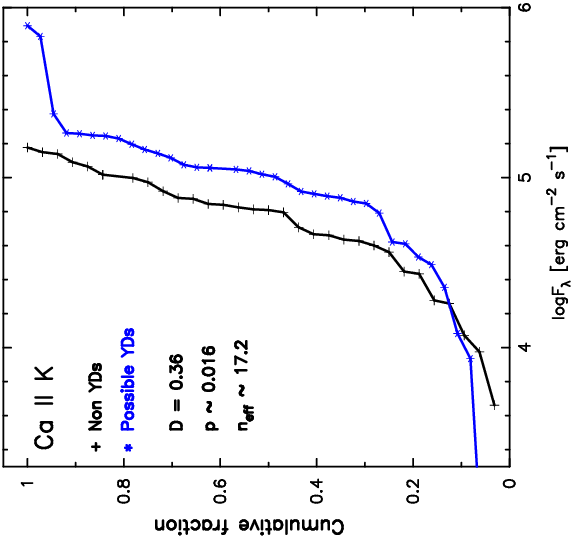}
\end{minipage}
\begin{minipage}{0.49\linewidth}
\includegraphics[angle=270,scale=0.45]{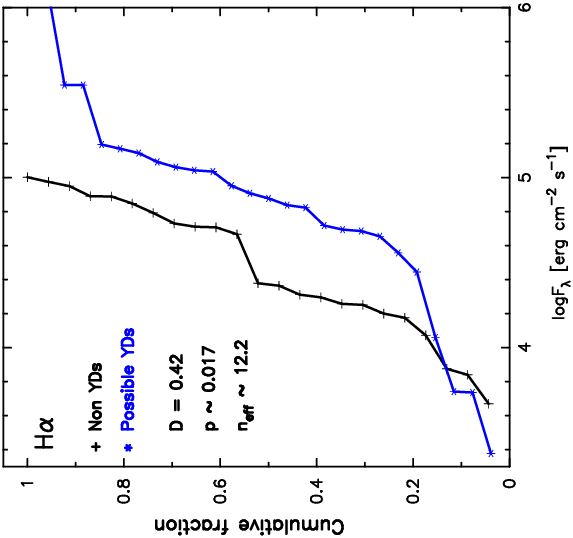}
\end{minipage}
\caption{
 Cumulative distribution function of $\log F_{\lambda}$ for
 the Ca~{\sc ii} K (left) and  H$\alpha$ (right) lines.}
\label{yd_old_comparison}
\end{figure}

  A total of 37 stars were classified as  kinematically ``young''(see Sec.~\ref{spatial_velocity_subsec}),
  although as cautioned some of them might indeed by old field stars.
  In order to  compare
  the $\log F_{\lambda}$ values between possible young and
  old stars, a series of two sided Kolmogorov-Smirnov (K-S) tests were performed.
  The results are given in Table~\ref{ksteststable}, while
  the cumulative distribution
  functions of $\log F_{\lambda}$ for two lines
  (Ca~{\sc ii} K and H$\alpha$) are shown in Figure~\ref{yd_old_comparison}.
  The results show  a clear tendency for kinematically young stars to show
  higher levels of activity in the Ca~{\sc ii} H \& K lines,
  as well as in the Balmer lines. 
  We are not able to reject the null
  hypothesis of both samples coming from the same parent distribution
  when considering the Balmer line  H$\gamma$  line but even
  for this line we should note the very low $p$-value returned by the 
  K-S analysis. 


\begin{table}
\centering
\caption{ 
Results of the K-S tests performed in this work between
possible young disc and old stars. We consider a
confidence level of 98\% in order to reject the null hypothesis H$_{0}$
(both samples coming from the same underlying continuous
distribution).
}
\label{ksteststable}
\begin{tabular}{lcccccc}
\hline\noalign{\smallskip}
Line  &  n$_{\rm old}$ & n$_{\rm young}$ & n$_{\rm eff}$ & $D$  & $p$   & H$_{0}^{\ddag}$ \\
\hline
Ca~{\sc ii} H	&	33	&	35	&	17	&	0.41  &	0.004  & 1	\\
Ca~{\sc ii} K	&	32	&	37	&	17	&	0.36  &	0.016  & 1	\\
H$\alpha$   	&	23	&	26	&	12	&	0.42  &	0.017  & 1	\\
H$\beta$  	&	25	&	31	&	14	&	0.45  &	0.004  & 1	\\
H$\gamma$   	&	22	&	31	&	13	&	0.35  &	0.069  & 0	\\
H$\delta$       &	25	&	30	&	14	&	0.42  &	0.011  & 1	\\
H$\epsilon$ 	&	33	&	35	&	17	&	0.36  &	0.019  & 1	\\
X-ray           &       12      &       17      &        7      &       0.29  & 0.500  & 0      \\  
\hline
\end{tabular}
\tablefoot{ 
 $D$ is the maximum deviation between the empirical distribution
 functions of samples 1 and 2; $p$ corresponds to the estimated
 likelihood of the null hypothesis, a  value that is known to be reasonably
 accurate for sample sizes for which n$_{\rm eff}$ $=$ (n$_{\rm 1}$ $\times$ n$_{\rm 2}$)/(n$_{\rm 1}$ + n$_{\rm 2}$) $\ge$ 4;
 H$_{0}^{\ddag}$ (0): Accept null hypothesis; (1): Reject null hypothesis.}
\end{table}

\subsubsection{X-ray emission versus stellar parameters}

 We now consider the relationships between rotation and stellar parameters
 and the X-ray emission. The comparison of the X-ray emission with
 the optical fluxes is presented in Section~\ref{chromos_corona_comp}.
 A total of 29 among our stars have available X-ray data.
 This figure represents $\sim$ 41\% of the total sample.
 The fraction of stars with X-ray detections is slightly smaller for the stars
 identified as binaries ($\sim$ 31\%) than the one for stars without known
 stellar companions ($\sim$ 47\%).

\begin{figure*}[!htb]
\centering
\begin{minipage}{0.30\linewidth}
\includegraphics[angle=270,scale=0.30]{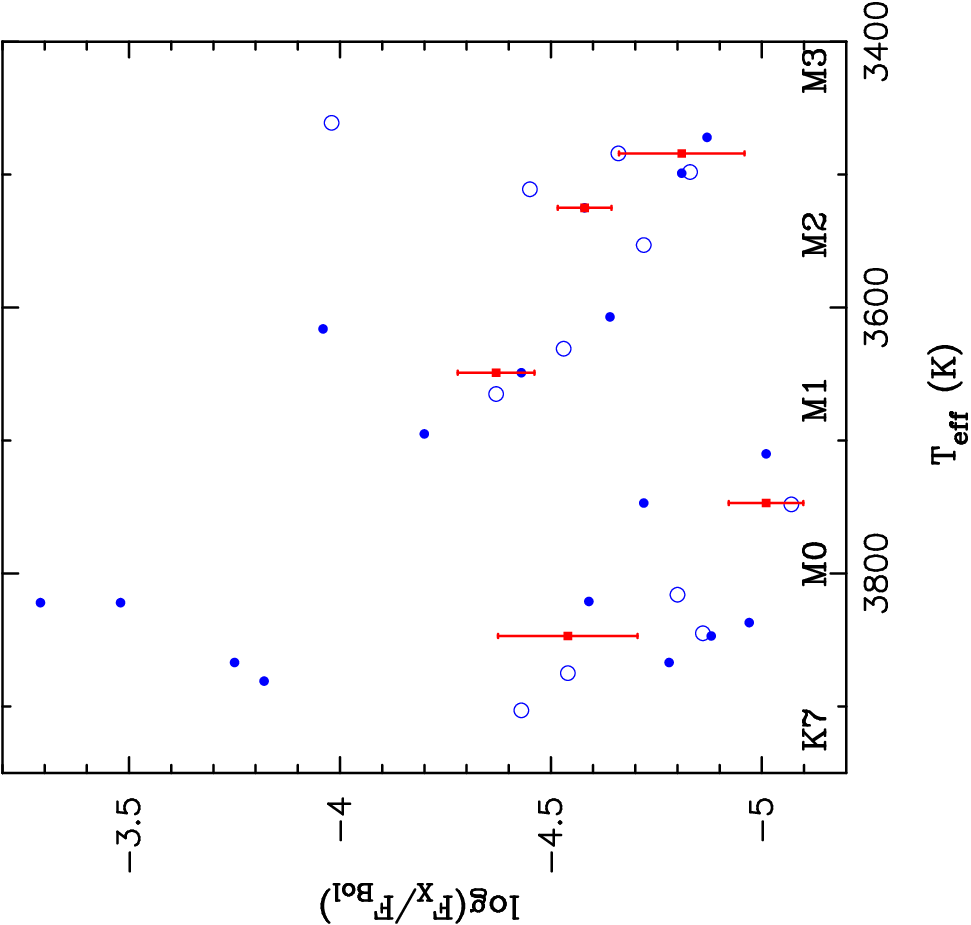}
\end{minipage}
\begin{minipage}{0.30\linewidth}
\includegraphics[angle=270,scale=0.30]{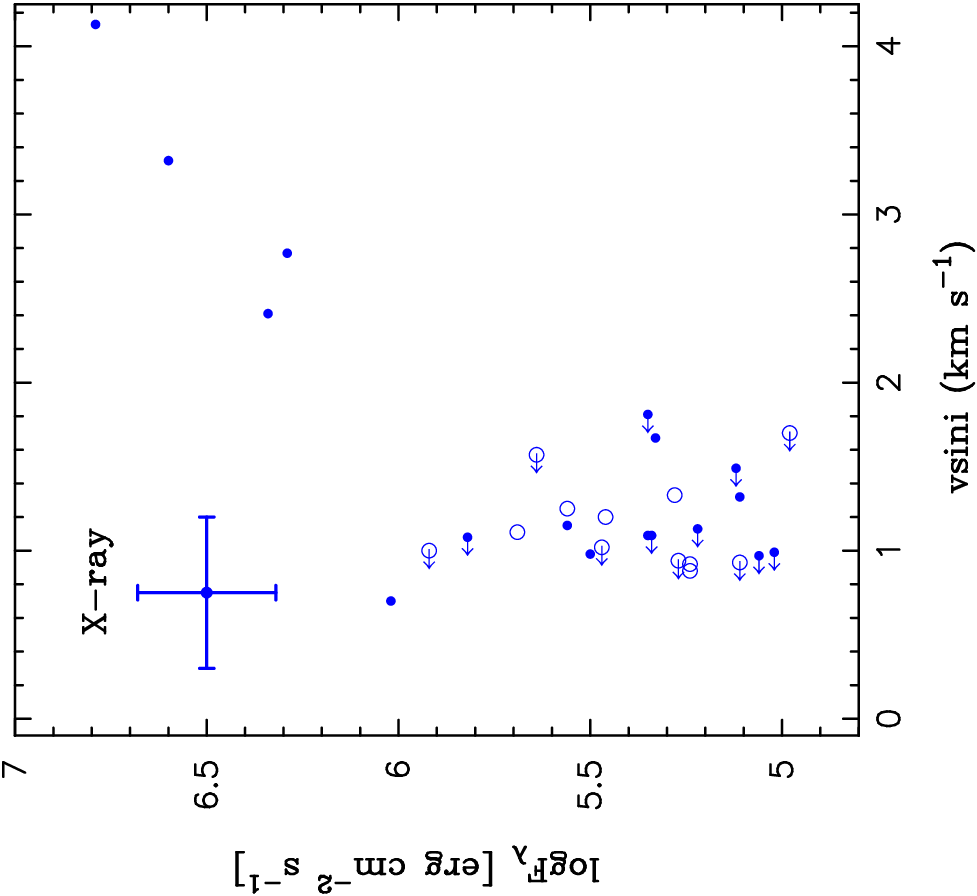}
\end{minipage}
\begin{minipage}{0.30\linewidth}
\includegraphics[angle=270,scale=0.30]{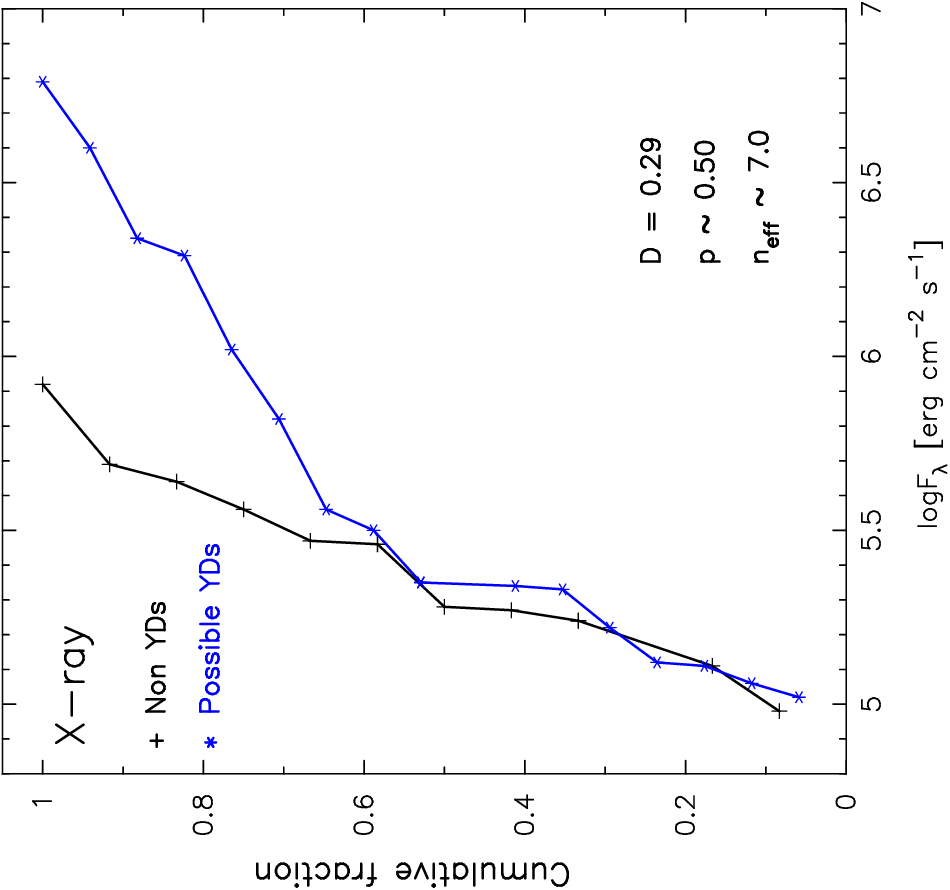}
\end{minipage}
\caption{
 X-ray emission as a function of the stellar parameters.
 Left panel: $\log F_{X}$/$\log F_{\rm Bol}$ vs. stellar effective temperature.
 Middle panel: X-ray flux, $\log F_{X}$, vs. v$\sin i$.
 Right panel: Cumulative distribution function of $\log F_{X}$.
 Colours and symbols are as in previous figures. }
\label{xray_flux_parameters}
\end{figure*}

 The analysis of the X-ray emission as a function of the effective
 temperature, Figure~\ref{xray_flux_parameters} left panel, shows a significant scatter 
 with $\log F_{\rm X}$/$\log F_{\rm Bol}$ values ranging from -3.3 to -5.1,
 although most of the stars ($\sim$ 79\%) show values between -4.25 and -5.1. 
 These values are lower than the median value of -3.95 found by \cite{2013MNRAS.431.2063S}
 in a study of the nearby (within 10 pc) M dwarfs, although this work includes
 M dwarfs up to spectral type M7. 
 The figure does not reveal a clear trend of the X-ray emission with the
 effective temperature in line with the results found when considering
 the optical activity indicators (Section~\ref{rotteff}).

 The middle panel in Figure~\ref{xray_flux_parameters} shows $\log F_{X}$ as a function
 of the projected rotational velocities v$\sin i$. 
 It can be seen that for low ($<$ 2 km s$^{\rm -1}$) v$\sin i$ values the scatter in
 the X-ray values is large. 
 The stars with the largest v$\sin i$ also show the largest X-ray emission.
 The statistical analysis of the data, see Table~\ref{asurv_statistics},
 shows that the probability that the X-ray fluxes and v$\sin i$ are correlated
 by chance is relatively low, $\sim$ 1-2\% with correlation coefficients of the
 order of 0.40. 

 We also compared the distribution of X-ray emission for the kinematically
 old and possible young stars, Figure~\ref{xray_flux_parameters} right panel.
 As a whole, there seems to be no difference between possible young and old
 stars (a K-S tests returns the values $D$ $\sim$ 0.29, $p$ $\sim$ 0.50).
 However, the figure reveals that while the $\log F_{\rm X}$ distribution
 of  kinematically selected young/old stars seems to be identical for values of X-ray emission
 $\log F_{\rm X}$ $<$ 5.5 erg cm$^{\rm -2}$ s$^{\rm -1}$, at larger values
 possible young stars clearly tend to show larger X-ray emission values than
 old stars. 
 
\subsection{Balmer decrements}

  Ratios between pair of fluxes, in particular the Balmer decrements
  (e.g. H${\alpha}$/H${\beta}$), are indicators of the physical
  conditions of the emitting regions \citep[e.g.][]{1979ApJ...230..581L,1991PhDT.........9C}. 

\begin{figure}
\centering
\includegraphics[angle=270,scale=0.45]{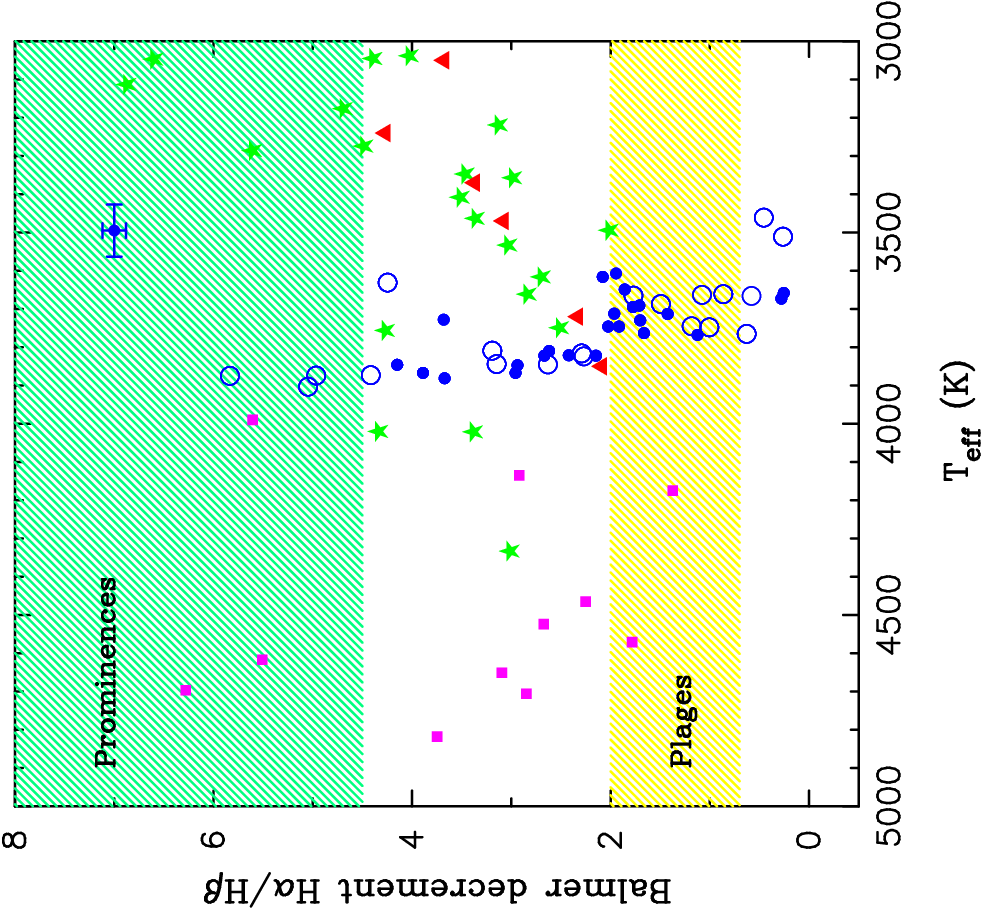}
\caption{
 Balmer decrement F$_{H\alpha}$/F$_{H\beta}$ 
 vs. the effective temperature. Possible young stars are shown with filled symbols.
 Green stars denote pre-MS M stars 
 from \cite{2013A&A...558A.141S}, red triangles correspond to the
 active M dwarf templates from the Sloan Digital Sky Survey \citep{2007AJ....133..531B},
 while data from \cite{2015A&A...575A...4F} is shown in purple squares.
 Typical ranges of solar plages and prominences are shown
 as hatched areas. Typical uncertainties are also shown.} 
\label{balmer_decrement}
\end{figure}

  Figure~\ref{balmer_decrement} shows the Balmer decrement F$_{H\alpha}$/F$_{H\beta}$ as a function
  of the effective temperature. Typical values of solar plages and prominences
  \citep[see e.g.][]{1979ApJ...230..581L,1991PhDT.........9C} are overplotted for comparison.
  The figure shows the large range of F$_{H\alpha}$/F$_{H\beta}$ values covered
  by our sample. 
  The stars in our sample with T$_{\rm eff}$ > 3800 K
  show a decreasing trend in F$_{H\alpha}$/F$_{H\beta}$ as we move towards cooler
  stars. Three of our hottest stars (GJ 9404, GJ 548A, and StKM 1-650) 
  show values of the Balmer decrement
  compatible or above the region of solar prominences\footnote{  
  The stars BPM 96441, and 2MASS J22353504+3712131 were excluded from the analysis
  given that their large errors in F$_{H\beta}$ 
  make the derived decrements unreliable.}. 
  For stars in the spectral range between $\sim$ M0 and M1.5 the Balmer decrements
  F$_{H\alpha}$/F$_{H\beta}$  most of them show 
  values compatible with those of solar plages.

  Our results can be compared with pre-MS M stars
  \citep{2013A&A...558A.141S} and
  the active M dwarf templates from the Sloan Digital Sky Survey
  \citep{2007AJ....133..531B},
   and the results by \citet{2015A&A...575A...4F} who analysed the Balmer decrement of magnetically active stars
  and accretors in two young regions (namely, Chamaeleon I and $\gamma$~Velorum), with the exception of two
  strong accreting stars in their sample, which have values of $F_{H\alpha}/F_{H\beta}$ close to
  $\sim$ 30. All these samples are also shown in Figure~\ref{balmer_decrement}.
   It is clear from the figure that both the pre-MS sample as well
  as the active M dwarf templates show a trend of increasing 
  F$_{H\alpha}$/F$_{H\beta}$ decrement with decreasing temperature.
  As this is somehow the opposite of what we found for our K7-M0 dwarfs
  one might speculate with the possibility of some bias affecting 
  our earliest stars as, for example, their slightly higher v$\sin i$ on average
  (as larger decrements might be expected for faster rotating stars).
  However, the values of the F$_{H\alpha}$/F$_{H\beta}$ decrements shown by 
  our stars with T$_{\rm eff}$ > 3800 K  are very well in agreement with the decrement
  values found in the pre-MS and in the \citet{2015A&A...575A...4F} samples
  (roughly between 2.0 and 5.0, i.e., between solar plages and prominences).
  For our M dwarfs with T$_{\rm eff}$ < 3800 K, the values of the Balmer decrements
  are clearly lower than the literature samples (below 2.0) confirming the
  low-activity levels of our sample. 

  Finally, we also note that the Balmer decrement show
  no difference between possible young
  and old disc stars except for the fact that there seems to be no 
  kinematically young stars among the
  ``prominence-like'' stars. 

\subsection{Flux-flux relationships}
  
\begin{figure*}[!htb]
\centering
\begin{minipage}{0.48\linewidth}
\includegraphics[angle=270,scale=0.45]{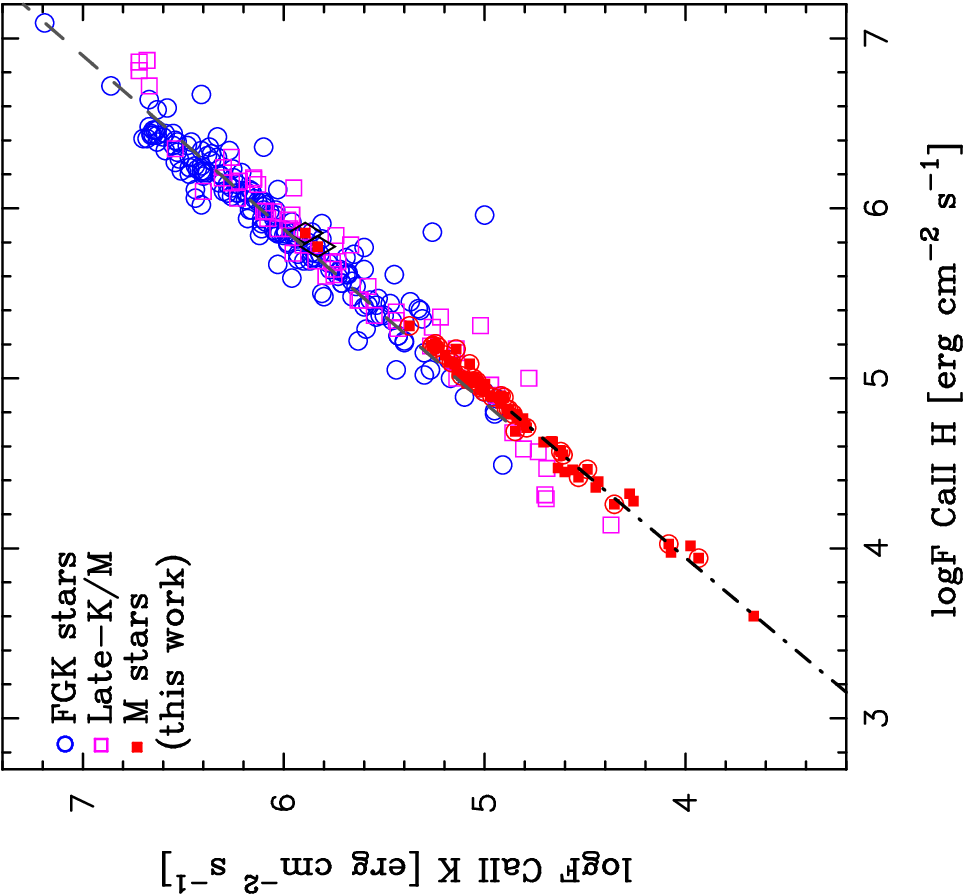}
\end{minipage}
\begin{minipage}{0.48\linewidth}
\includegraphics[angle=270,scale=0.45]{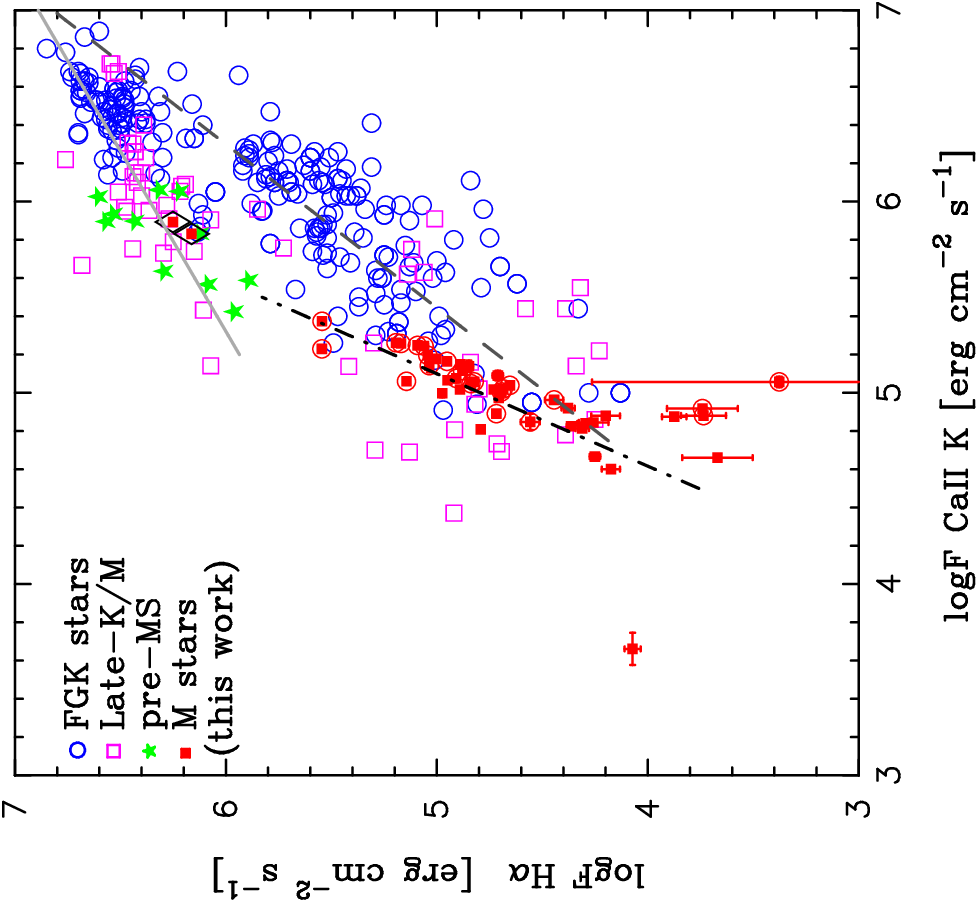}
\end{minipage}
\caption{
Flux-flux relationships between calcium lines (Ca~{\sc ii} H \& K, left panel),
and between H${\alpha}$ and Ca~{\sc ii} K (right panel).
M dwarfs from this work are plotted with red filled squares; FGK stars from 
\cite{2010A&A...514A..97L,2010A&A...520A..79M,2011MNRAS.414.2629M} 
with open circles, late-K and M stars from
the literature (same references than for FGK stars) are shown in purple open squares; 
green stars denote the M0-M3 pre-MS M stars from \cite{2013A&A...558A.141S}.
Possible young disc stars in our M star sample are shown with circles.
The two stars discussed in Sec.~\ref{ofinterest} are indicated with diamonds.
The dot-to-dash black line represents our best fit; the relations for the
``active'' and ``inactive'' branches by \cite{2011MNRAS.414.2629M} are shown
in light grey solid and  dashed dark grey lines respectively.
}
\label{flux_flux_relationships}
\end{figure*}

\begin{figure*}[!htb]
\centering
\begin{minipage}{0.48\linewidth}
\includegraphics[angle=270,scale=0.45]{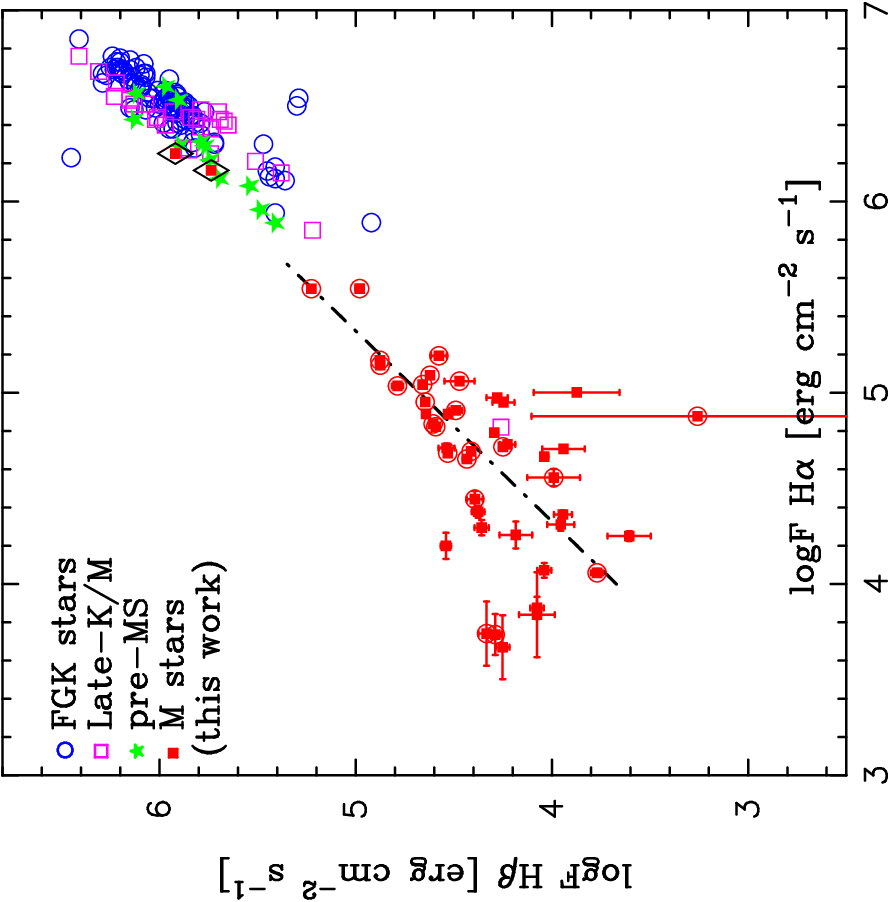}
\end{minipage}
\begin{minipage}{0.48\linewidth}
\includegraphics[angle=270,scale=0.45]{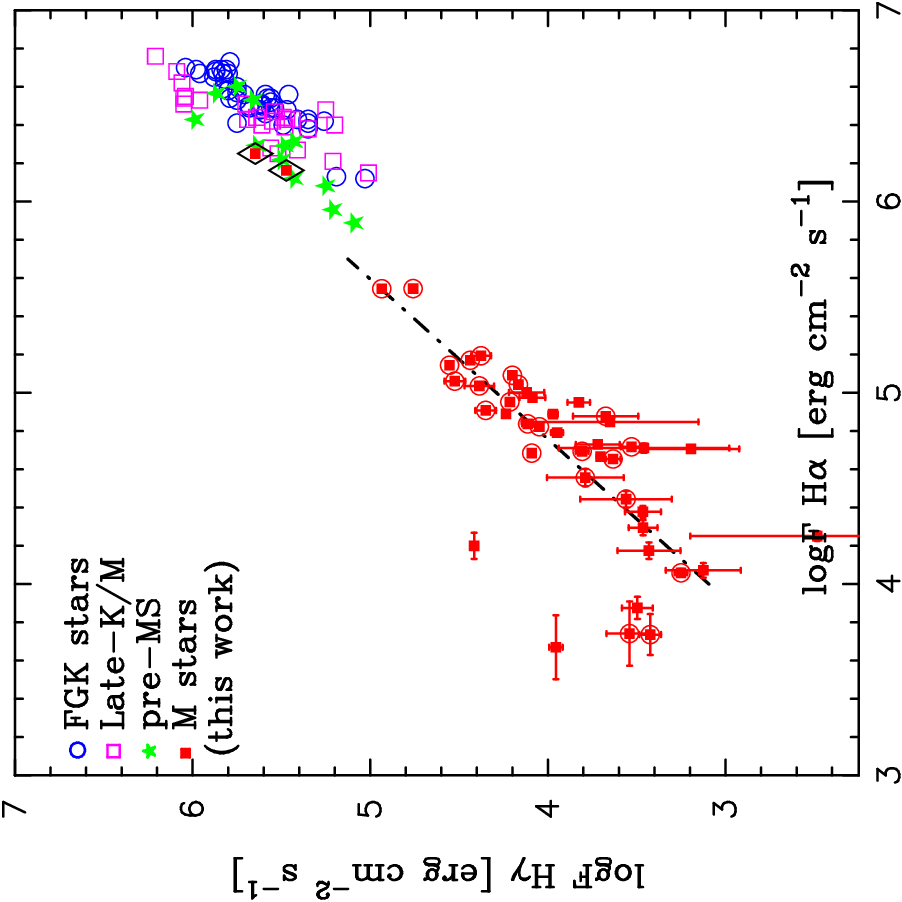}
\end{minipage}
\begin{minipage}{0.48\linewidth}
\includegraphics[angle=270,scale=0.45]{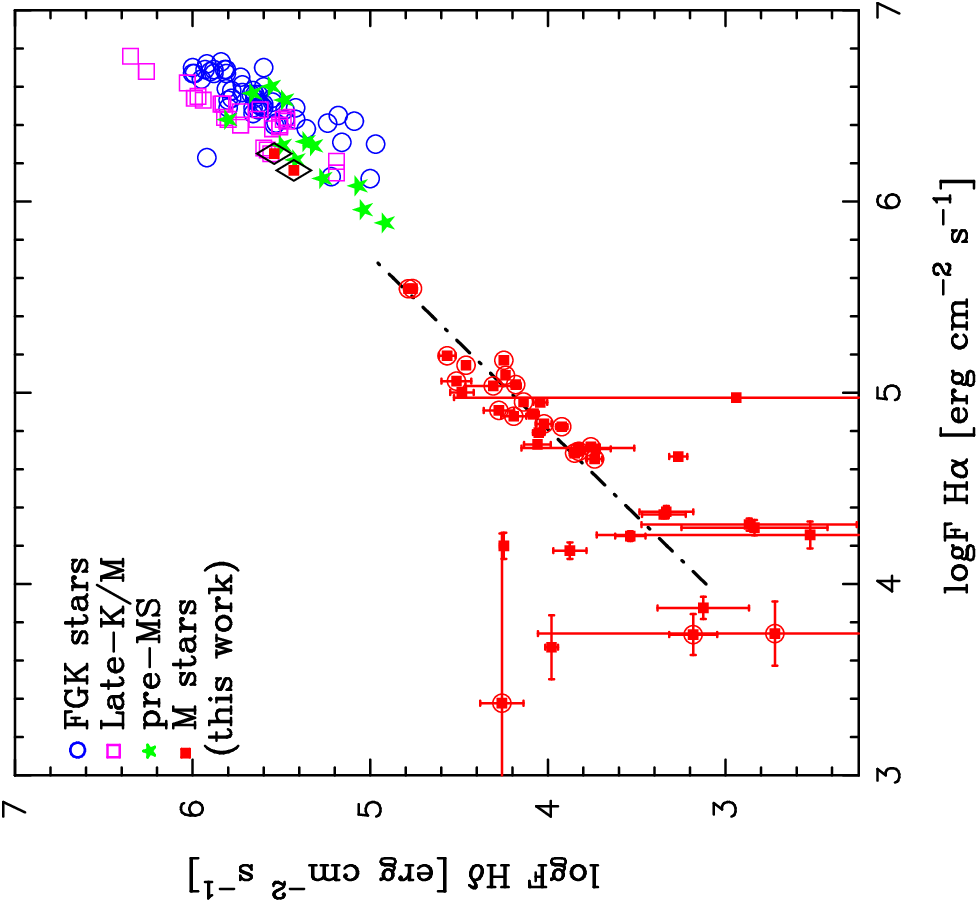}
\end{minipage}
\begin{minipage}{0.48\linewidth}
\includegraphics[angle=270,scale=0.45]{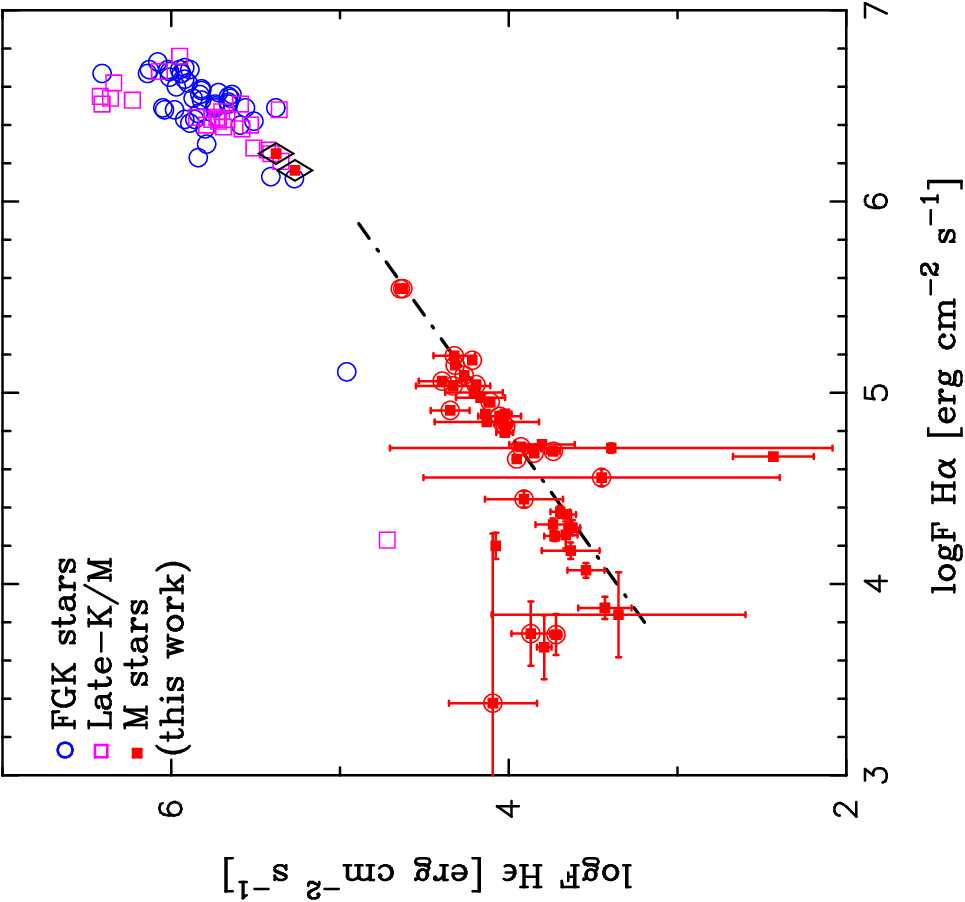}
\end{minipage}
\caption{
Flux-flux relationships
between Balmer lines: 
H${\alpha}$, H${\beta}$, (top left panel)
H${\alpha}$, H${\gamma}$ (top right panel),
H${\alpha}$, H${\delta}$ (bottom left), and
H${\alpha}$, H${\epsilon}$ (bottom right).
Colours and symbols are as in Figure~\ref{flux_flux_relationships}.
}
\label{flux_flux_relationships2}
\end{figure*}

  Figures~\ref{flux_flux_relationships} and ~\ref{flux_flux_relationships2}
  show the comparison between
  pairs of fluxes of different chromospheric lines for the stars in our sample.
  Power-law functions were fitted to the data: 

\begin{equation}
 \log F_{\rm 1} = a_{\rm 0} + a_{\rm 1} \log F_{\rm 2}
\end{equation}
 
  \noindent where F$_{\rm 1}$ and F$_{\rm 2}$ are the fluxes of
  two different lines and a$_{\rm 0}$ and a$_{\rm 1}$ the fit
  coefficients. We recall here that we are considering the flux excesses
  measured over the combined spectra, so for each star all the fluxes are obtained
  from the same average spectrum. 
  Several samples are overplotted for comparison:
  a sample of F, G, and K stars from \cite{2010A&A...514A..97L,2010A&A...520A..79M,2011MNRAS.414.2629M};
  a sample of late-K and M dwarfs (from the same authors); and a sample
  of pre-MS M stars from \cite{2013A&A...558A.141S}.
  For better comparison with our data, from the latest sample only stars in the spectral range K7-M3 were
  considered.
  The values of a$_{\rm 0}$ and a$_{\rm 1}$ are given in
  Table~\ref{flux_flux_coeff}. 
  The fits were performed with the least-squares bisector regression described
  by \cite{1990ApJ...364..104I}. 
  Stars with large errors in the fluxes were excluded from the fits. 


  Our sample of M dwarfs seems to follow the same trend as
  FGK stars and other late-K/early-M dwarfs  in the Ca~{\sc ii} H vs
  Ca~{\sc ii} K plane (Figure~\ref{flux_flux_relationships}, left panel)
  without any obvious deviation between both samples.
  The H${\alpha}$ vs. Ca~{\sc ii} K plot is shown in the right panel
  of Figure~\ref{flux_flux_relationships}.
  \cite{2011MNRAS.414.2629M} identified two branches in the flux-flux relationships  
  when one of the considered activity diagnostics is the H${\alpha}$ line.
  The ``inactive'' branch is composed by the majority of the stars and occupied
  by field stars with spectral types from F to M. The deviating stars, on the
  other hand, constitute the upper or ``active'' branch 
  which is composed of young late-K and M dwarfs with saturated
  H${\alpha}$ emission.  
  In Figure~\ref{flux_flux_relationships}, right panel, 
  it can be seen that our M dwarfs are located in the region of the plot corresponding
  to the inactive branch. However, our derived slope is 
   steeper ($\sim$ 2)  than the previously reported values ($\sim$ 1).
  In addition, a vertical offset seems to be present in our sample
  when comparing with the literature sample.
  It seems that as a whole, our sample of M dwarfs is located slightly above the
  inactive branch, or in other words, there seems to be a lack of stars with low H$\alpha$ emission.

  We also note that two of our targets (namely TYC3720-426-1, TYC2703-706-1) seem to follow the same
  tendency of the  stars in the active branch. These two stars were not considered in the
  fits and are discussed in more details in the next subsections.

\begin{table*}
\centering
\caption{
Coefficients of the flux-flux relationships. 
 }
\label{flux_flux_coeff}
\begin{tabular}{llccl}
\hline\noalign{\smallskip}
\multicolumn{2}{c}{}                   &  \multicolumn{2}{c}{This work}              & Other works \\                          
$\log$F$_{\rm 1}$ & $\log$F$_{\rm 2}$  &   a$_{\rm 0}$                     & a$_{\rm 1}$         & a$_{\rm 1}$ \\
\hline
Ca~{\sc ii}, K & Ca~{\sc ii}, H & 0.0121  $\pm$  0.0006   & 1.0110 $\pm$ 0.0001  &  0.99 $\pm$ 0.03$^{(d)}$,  0.98 $\pm$ 0.02$^{(e)}$, 
0.86 $\pm$ 0.08$^{(f)}$  \\
H$\alpha$      & Ca~{\sc ii}, K &  -5.55 $\pm$  0.02   & 2.069 $\pm$ 0.003    &
1.13 $\pm$ 0.10$^{(a,c)}$, 0.95 $\pm$ 0.08$^{(d)}$,  1.20 $\pm$ 0.07$^{(e)}$, 1.26 $\pm$ 0.15$^{(f)}$, 0.69 $\pm$ 0.08$^{(g)}$ \\
\hline
H$\beta$       & H$\alpha$ &  -0.345  $\pm$ 0.008  & 1.004 $\pm$  0.002   & 1.03 $\pm$ 0.07$^{(f)}$, 1.19 $\pm$ 0.09$^{(g)}$ \\
H$\gamma$      & H$\alpha$ &  -1.689  $\pm$ 0.010  & 1.196 $\pm$  0.002   & 1.06 $\pm$ 0.05$^{(f)}$, 1.18 $\pm$ 0.11$^{(g)}$ \\
H$\delta$      & H$\alpha$ &  -1.29   $\pm$ 0.02   & 1.101 $\pm$  0.003   & 1.20 $\pm$ 0.08$^{(f)}$, 1.23 $\pm$ 0.16$^{(g)}$ \\
H$\epsilon$    & H$\alpha$ &   0.11   $\pm$ 0.01   & 0.812 $\pm$  0.002   & 0.79 $\pm$ 0.10$^{(a,c)}$,  2.96 $\pm$ 0.87$^{(f)}$   \\  
\hline
F$_{\rm X}$ & Ca~{\sc ii}, K & 1.22  $\pm$ 0.14  & 0.86  $\pm$ 0.03  & 2.38 $\pm$ 0.26$^{(b,c)}$, 1.06 $\pm$ 0.17$^{(g)}$ \\
F$_{\rm X}$ & H$\alpha$      & 1.95  $\pm$ 0.15  & 0.76  $\pm$ 0.03  &
2.11 $\pm$0.20$^{(a,c)}$, 1.60 $\pm$ 0.07$^{(e)}$,  1.29$\pm$0.15 $^{(f)}$, 1.50 $\pm$ 0.14$^{(g)}$, 1.89 $\pm$ 0.31$^{(h)}$  \\
\noalign{\smallskip}\hline\noalign{\smallskip}
\end{tabular}
\tablefoot{
$^{(a)}$ \cite{1995A&A...294..165M};  
$^{(b)}$ \cite{1996A&A...312..221M};  
$^{(c)}$ \cite{1996ASPC..109..657M};  
$^{(d)}$ \cite{2010A&A...520A..79M};  
$^{(e)}$ \cite{2011MNRAS.414.2629M};  
$^{(f)}$ \cite{2012A&A...537A..94S};  
$^{(g)}$ \cite{2013A&A...558A.141S};  
$^{(h)}$ \cite{2013MNRAS.431.2063S}.  
}
\end{table*}

  Regarding the slopes between different Balmer lines,
  Figure~\ref{flux_flux_relationships2} shows that
  our sample of M dwarfs 
  follows a similar tendency as the solar-type stars from the literature. %
  The sample of pre-MS dwarfs
  also behave
  in the same way as the rest of stars.
  Our derived slopes for the flux-flux relationships have values
  $\sim$ 1.0/1.2 and are compatible with previously reported values
  (even though these literature values correspond to studies of 
  pre-MS stars and/or  include ``ultra-cool'' dwarfs, see references in Table~\ref{flux_flux_coeff}).
  The H$\epsilon$ line shows a slightly lower slope than 1.0 ($\sim$ 0.80),
  in agreement with previous works \citep{1995A&A...294..165M,1996ASPC..109..657M},
  although for ultra-cool dwarfs (including stars up to spectral type M9) \cite{2012A&A...537A..94S} reports a slope
  close to 3.0.

   We should caution that there are less literature stars to compare with
  as the Balmer fluxes for a large fraction of the literature stars are not
  reported. Further, nearly all stars in the comparison samples with
  available fluxes in the Balmer lines are stars 
  of the ``active'' branch. This explains the apparent lack of
  comparison stars at low fluxes in Figure~\ref{flux_flux_relationships2}
  (in contrast to the $\log F$(H$\alpha$) vs. $\log F$(Ca~{\sc ii} K) plot, 
  Figure~\ref{flux_flux_relationships}, where our M 
  stars are mixed with the literature M dwarf sample).

  We conclude that our sample of M dwarfs is complementary 
  to the literature samples in the sense that they follow similar
  flux-flux relationships. Our sample constitutes an ``extension'' of  
  the analysis of the Ca~{\sc ii} H \& K and Balmer flux-flux relationships
  of main-sequence M dwarfs to the very low flux domain. 

\subsection{Chromospheric-corona flux-flux relationships}
\label{chromos_corona_comp}

\begin{figure*} 
\centering
\begin{minipage}{0.48\linewidth}
\includegraphics[angle=270,scale=0.45]{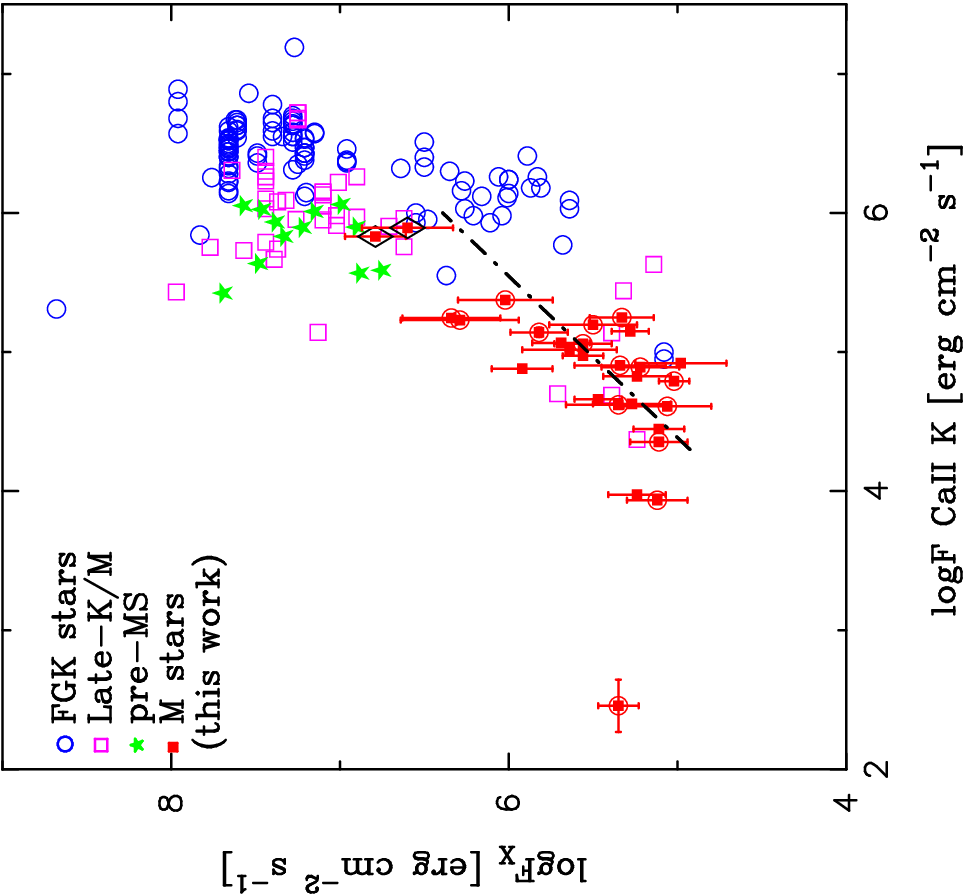}
\end{minipage}
\begin{minipage}{0.48\linewidth}
\includegraphics[angle=270,scale=0.45]{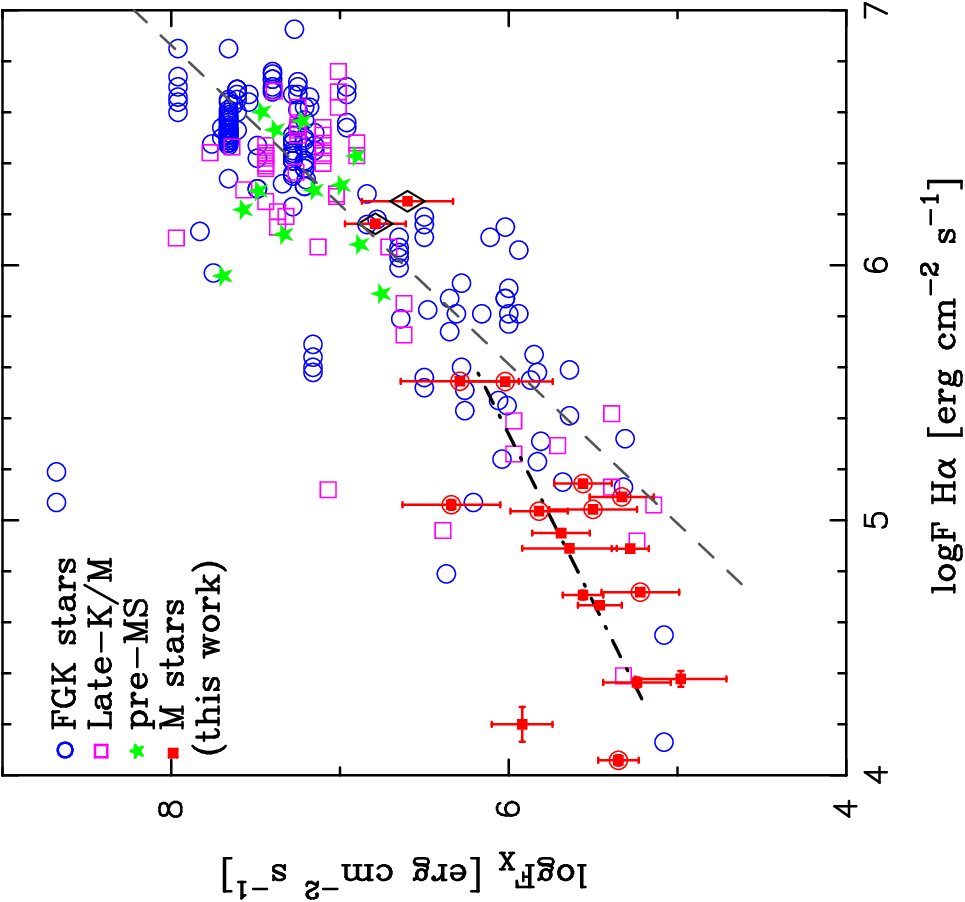}
\end{minipage}
\caption{
Flux-flux relationships between X-ray 
and the calcium line Ca~{\sc ii} K  (left panel),
and between X-ray and the H${\alpha}$ line (right panel).
Colours and symbols are as in Figure~\ref{flux_flux_relationships}.}
\label{flux_x}
\end{figure*}

 In addition to the flux-flux relationships between different chromospheric
 activity indicators, the chromospheric-coronal relation was also studied.
 All stars with X-ray detections show Ca~{\sc ii} K emission with only
 one exception (GJ 412A), regarding H$\alpha$ only 69\% of the stars
 with X-ray data show emission in this line.
 Figure~\ref{flux_x} shows the X-ray flux,
 $\log F_{\rm X}$, as a function of the
 fluxes in the Ca~{\sc ii} K line (left) and H$\alpha$ (right).
 

  No distinction between ``active'' and ''inactive'' branches
  was found in the literature  for the $\log F_{\rm X}$ vs.
  Ca~{\sc ii} K line and  H$\alpha$  flux-flux relationships
  \citep[][]{2011MNRAS.414.2629M}\footnote{The authors, however, do find
  the two branches in the X-ray vs. Ca~{\sc ii} IRT, 8498\AA \space line
  analysis. Unfortunately, the Ca~{\sc ii} IRT lines are not covered by
  our spectra.}.
  Our sample of M dwarfs seems
  to follow the same tendency than the literature estimates.
  Our analysis, however, clearly reveals that our M dwarfs have 
   lower levels of X-ray fluxes than the FGK stars.
  Figure~\ref{flux_x} also shows those of our M dwarfs with
  larger ``deviations'' from the literature samples 
   are the ones with lowest X-ray and 
  chromospheric emission.

  Our lower levels of X-ray fluxes translates into significant lower slopes 
  than the ones
  previously reported in the literature (see Table~\ref{flux_flux_coeff}).
  As before, it is important to note that some previous works
  are based on pre-MS or include cooler stars than ours.
  Also, the analysis presented in 
  \citet{1995A&A...294..165M,1996ASPC..109..657M}
  is based on binary stars in chromospherically active systems
  which might explain their significantly higher ($>$ 2.0) slopes.
  
  The two stars discussed in Section~\ref{ofinterest} 
  are in the same place or close to the place occupied by the pre-MS M stars
  in Figure~\ref{flux_x}. Further, they show
  levels of X-ray activity compatible or close to saturation (L$_{\rm X}$/L$_{\rm Bol}$ $\sim$ 10$^{\rm -3}$)
  and were identified in the ``active'' branch in the
  H$\alpha$ vs. Ca~{\sc ii} K plot. This agrees with previous works suggesting that the stars in the ``active'' branch
  are young or flare stars with saturated X-ray emission \citep[e.g.][]{2011MNRAS.414.2629M}.

  
\subsection{Notes on interesting stars}\label{ofinterest}

 Here, we give further details on the two stars the position of which in 
 the flux-flux diagrams is consistent with the position of the 
 1-10 Myr old pre-MS stars from the literature. 

\subsubsection*{TYC3720-426-1}
  This M0 star has the largest v$\sin i$ (4.07 kms$^{\rm -1}$) and
  strongest X-ray emission ($\log$(L$_{\rm X}$/L$_{\rm Bol}$) = -3.29)
  in our sample. 
  It shows emission in all the Balmer lines and in the Na~{\sc i} D$_{\rm 1}$, D$_{\rm 2}$ doublet.
  It is the second star in our sample with the largest median S index.
  Its kinematics are compatible with being a member of the Local Association
  of stars, in particular \cite{2011ApJ...732...61Z,2013ApJ...762...88M} classified it as a member of the Columba
  nearby young association ($\sim$ 30 Myr). 
  Using the BANYAN II on-line tool \citep{2013ApJ...762...88M,2014ApJ...783..121G}, which implements a Bayesian analysis to determine 
  probabilities of membership to nearby young kinematic groups,
  together with the HADES radial velocities, we found a 61\% membership probability
  for this star to be a member of the Columbia association. 
  From the SuperWASP archive\footnote{http://exoplanetarchive.ipac.caltech.edu/docs/SuperWASPMission.html}
  photometry a rotation period of 4.65 $\pm$ 0.03 days can be inferred. 
  The position of this star in a period-colour diagram agrees reasonably well with the age
  of the Columba association. 

\subsubsection*{TYC2703-706-1}
  TYC2703-706-1 is an M0.5 dwarf which has the largest 
  activity S index in our sample. All its activity indicators
  (including the Na~{\sc i} D$_{\rm 1}$, D$_{\rm 2}$ doublet and the He~{\sc i} D$_{\rm 3}$ line)
  appear in emission. It also has one of the highest rotation levels
  (v$\sin i$ of  3.32 kms$^{\rm -1}$) and a large fractional X-ray 
  luminosity ($\log$(L$_{\rm X}$/L$_{\rm Bol}$) of -3.48). Its galactic
  spatial velocity components suggest that it is a young disc star.
  Indeed, it has been identified as a candidate member\footnote{Flagged with a value of 2 in a scale of 1 to 4, where 4 means
  the best candidates.}
  of the young ($\sim$ 23 Myr)
  $\beta$ Pic stellar association \citep{2012AJ....143...80S}. However,
  our analysis with the BANYAN II tool reports a probability of 0\%
  of being a member of this association.
  The analysis of the SuperWASP photometry reveals a rotation period
  of 8.00 $\pm$ 0.05 days. This period falls in the upper boundary of the period distribution
  of the bona fide $\beta$ Pic members, suggesting an age
  equal or slightly younger than the $\beta$ Pic members.

 
\section{Discussion and conclusions}
\label{summary}

  In spite of the increasing effort devoted to study M dwarfs,
  our understanding of their chromospheres and the processes that we
  generally call ``activity'' are still very far from being fully
  understood. In this work, a detailed analysis of the relationships between
  activity and stellar parameters (rotational velocity, effective temperature, age)
  and the flux-flux relationships in a large sample of early M-dwarfs is presented. 
  Projected rotational velocities v$\sin i$ are computed using the
  CCF technique while emission excess fluxes in the Ca~{\sc ii} H \& K and
  Balmer lines are derived using the spectral subtraction technique.

  Besides the presence of some potential biases due to the fact that our
  sample was selected for a radial velocity search program (i.e., selected
  with low levels of activity) our study reveals several interesting trends:
  {\it i)} The strength of the chromospheric line emission seems to be constant in the spectral
  range studied here (M0-M3). This holds for all the activity indicators considered;
  {\it ii)} Field early M-dwarfs have very low v$\sin i$ values with a tendency of lower rotation levels as
  we move towards cooler stars up to spectral type M3, although this might need confirmation
  as the fraction of possible young stars seems to decrease towards cooler stars
  (there could also be a possible inversion for later spectral types as discussed in Sec.~\ref{rotteff});
  {\it iii)} The analysis suggest that a moderate but statistically significant
  correlation between activity and rotation might be present in our data; and
  {\it iv)}  Possible young stars show higher levels of emission 
  excess in the Ca~{\sc ii} H \& K lines and most of the Balmer lines than probable old disc stars.
  These findings agree reasonably well with previous works, suggesting that
  the already established activity-rotation-age relationship found in FGK stars also
  holds for early M-dwarfs.  

  The comparison between pairs of fluxes of different chromospheric lines
  has revealed other interesting trends.
  The analysis of the Balmer F$_{H\alpha}$/F$_{H\beta}$ decrement shows
  a trend of decreasing values, from values compatible with solar 
  prominences for stars with T$_{\rm eff}$ $\sim$ 3900 K, to values
  similar to those of the solar plages for T$_{\rm eff}$ $\sim$ 3750 K.
  Then, the Balmer decrement remains roughly constant in the range
  3750-3600 K.
 
  On the other hand, the analysis of the flux-flux relationships 
  shows that our M dwarfs sample is complementary to other 
  literature samples, extending the analysis of the flux-flux relationships
  to the low-chromospheric fluxes domain. 
  Our results confirm that field stars deviating from 
  the ``general'' flux-flux relationships are likely to be young.
  The low values of the chromospheric excess of our M stars
  is also revealed in the corona-chromosphere flux-flux
  relationships.
   We conclude that our sample represents a benchmark for the
  characterisation of magnetic activity at low levels.  


  Understanding the chromospheres of M-dwarfs is crucial for
  ongoing exo-planet searches. While first surveys tried to avoid
  ``active'' M dwarfs, there is ongoing evidence that M dwarfs
  show radial velocity signals due to the simultaneous presence
  of low-mass planets and activity-related phenomena
  \citep[e.g.][]{Affer}.
  Furthermore,
  thanks to new instrumentation at IR wavelengths where lines are less affected by magnetic activity
  than in the optical
  \citep[e.g.][]{2013hsa7.conf..842A,Carleo2016}
  new surveys are focusing on late-M and young stars.
  It is therefore mandatory to  understand the mechanisms involved
  in chromospheric and coronal heating
  as well as their dependence on the stellar parameters before
  a full understanding of their effects on exoplanet detection is reached.


\begin{acknowledgements}

 This work was supported by the Italian Ministry of Education,
 University, and Research  through the
 \emph{PREMIALE WOW 2013} research project under grant
 \emph{Ricerca di pianeti intorno a stelle di piccola massa}.
 GAPS acknowledges support from INAF through the \emph{Progetti Premiali} funding
 scheme of the Italian Ministry of Education, University, and Research.
 I. R. and M. P. acknowledge support from the Spanish
 Ministry of Economy and Competitiveness (MINECO) through grant ESP2014-57495-C2-2-R.
 J.I. GH. acknowledges financial support from the Spanish Ministry of Economy and Competitiveness (MINECO) under
 the 2013 Ram\'on y Cajal program MINECO RYC-2013-14875. 
 A. SM., J.I. GH., and R. R. also acknowledge financial support from the
 Spanish ministry project MINECO AYA2014-56359-P.
 G. M., L. A., E. M., G. P., and A. S. acknowledge support from the
 Etaearth project. 
 G. S., and I. P., acknowledge financial support from \lq\lq Accordo ASI--INAF\rq\rq\ n. 2013-016-R.0 Jul, 9 2013.
 A. Bayo and B. Montesinos are acknowledged for useful discussions
 about the BT-Settl models and interpolation. 

\end{acknowledgements}


\bibliographystyle{aa}
\bibliography{Mactivity.bib}

\Online
\section*{Online material}
 
  Table~\ref{parameters_table_full} lists all the stars analysed in this work.
  The table provides
  star identifier (Col. 1),
  effective temperature in Kelvin (Col. 2),
  spectral type (Col. 3),
  stellar metallicity in dex (Col. 4),
  stellar mass in solar units (Col. 5),
  stellar radius in solar units (Col. 6),
  logarithm of the surface gravity, $\log g$, in cms$^{\rm -2}$ (Col. 7),
  stellar luminosity, $\log (L_{\star}/L_{\odot})$ (Col. 8),
  and projected rotational velocities, $v \sin i$ in kms$^{\rm -1}$(Col. 9). 
  Each measured quantity is accompanied by its corresponding uncertainty.
 
  Table~\ref{kinematic_catalogue} gives the position and kinematic data:
  star identifier (Col. 1),  
  right ascension and declination (ICRSJ2000) (Cols. 2 and 3),
  proper motions in right ascension and declination
  in arcsec yr$^{\rm -1}$ (Cols. 4 and 5),
  stellar parallax with its uncertainty and reference (arcsec, Cols. 6),
  radial velocity in kms$^{\rm -1}$ (Col. 7), galactic spatial-velocity components $(U,V,W)$
  in kms$^{\rm -1}$
  (Cols. 8 - 10), and notes on binarity and possible membership
  to the young disc population. 

  Table~\ref{parameters_emission_excesses} provides the derived
  line emission excess fluxes, $\log$ F$_{\lambda}$ [erg cm$^{\rm -2}$ s$^{\rm -1}$],
  in the different chromospheric indicators studied in this work
  and X-ray. 

\end{document}